\documentclass[a4paper,10pt]{article}
\usepackage{ucs}
\usepackage[utf8]{inputenc}
\usepackage{amsmath}
\usepackage{amsfonts}
\usepackage{amssymb}
\usepackage{amsthm}
\usepackage{rotating}
\usepackage{xcolor}
\usepackage[english]{babel}
\usepackage[T1]{fontenc}
\usepackage{graphicx}
\usepackage{hyperref}
\usepackage{cuted}
\usepackage{verbatim}
\usepackage{subfigure}
\usepackage{multirow}
\usepackage{bbold}
\usepackage{slashed}
\usepackage{indentfirst}
\usepackage{tcolorbox}
\usepackage{placeins}
\usepackage[a4paper, total={7in, 10in}]{geometry}
\usepackage{setspace} 
\usepackage{physics}
\usepackage{float}
\usepackage{caption}

\DeclareMathAlphabet\mathbfcal{OMS}{cmsy}{b}{n}

\renewcommand{\d}{\mathcal{D}}

\newcounter{eq}

\begin{document}

\title{ \bf  
Large quark mass Euler-Heisenberg type action for QCD
  with quark sources
and its Abelian limit
}

\date{}

\author{ Igor de M. Froldi, Fabio L. Braghin,
% $^{1,2}$
\\
%\affiliation{$^1$
{\normalsize Instituto de F\'\i sica, Federal University of Goias,}
\\
{\normalsize Av. Esperan\c ca, s/n,
 74690-900, Goi\^ania, GO, Brazil}
}

\maketitle

\begin{abstract}
The quark determinant in the presence of a background gluon field is calculated in a large quark mass approach within a derivative expansion by considering quark sources. 
The resulting low-energy QCD Effective Field Theory (EFT) is valid for non-homogeneous fields and it
presents, besides the usual Euler-Heisenberg type gluon self-interactions, different types of corrections to the minimal quark-gluon interaction among which there are several imaginary terms.
The Abelian  limit of this effective action, i.e. an expanded Euler-Heisenberg effective action with fermion sources, is
also found and it is shown to 
contain, therefore, corrections to the fermion-photon interactions, including  imaginary terms.
Different resulting properties 
are calculated such as: their contributions to the  energy-momentum tensor and  its  trace, the Poynting vector,
 the  chromo-electric permittivity and chromo-magnetic permeability.
Conditions for these quantities, as well as for the corresponding Abelian limits, to be real are exhibited.
\end{abstract}

\section{Introduction }

The first historical example of a 
 low energy effective field theory (EFT) was found by Euler and Heisenberg
\cite{HeisenbergEulerOriginal,EulerHeisenberg1935}
by calculating the fermion 
 polarization contribution for the electromagnetic field dynamics.
The Euler-Heisenberg  (EH) effective action,  re-derived later by using different methods, 
 among others by Schwinger \cite{Schwinger1951},
has been extensively investigated
since then,
and it
is suitable 
 to shed light on different types of 
phenomena such as light-light scattering 
 birefringence
 and pair production 
\cite{DUNNE2005,lightlight,birefringence}.
Usually, its calculation reduces to the case of a constant electromagnetic field
being that only a  few different situations
 have been successfully reached 
 for time-dependent or spatially non-homogeneous configurations \cite{DUNNE2005,Karbstein2021}.
The standard one loop EH is one particle irreducible (1PI) although further loop corrections are not
\cite{Dunne-etal-2006,gies-karbstein,schubert-etal}.
Whereas in the case of the electromagnetic effective action,
the integration of the fermion field in the presence of background Abelian gauge fields  can be handled exactly, 
the equivalent effective action for non-Abelian   QCD, that 
 may include not only quark polarization but 
also gluon and ghost loops, usually
can be calculated  
by resorting  to some approximation method or in specific
limits \cite{Nielsen1978,Leutwyler1981,Schanbacher1982,Dittrich19831,Cea1988,Savvidy1977,NIELSEN1978304,gies, Ozaki2014,Galilo2011,EH-strongB,Ozaki2015a}.
  As an EFT, the QCD-EH effective action can be expected to provide reliable information 
concerning the low energy structure of QCD
and problems related to hadron structure. 
This effective action, so far, has been exploited 
in the absence of quark sources.
As a low energy EFT, it aims to describe corresponding 
effects and processes  within the range of validity of its derivation,
that may not easily addressed in terms of the fundamental theory.
An additional advantadge in the case of considering an EFT with physical degrees of freedom, as it is also the case of the EH effective action in eletromagnetism, is that it may contribute for
the identification or description of effective interactions usually inspired by phenomenology built with experimental results
whenever available. 
In the absence of 
specific experimental results, however, it may happen that
the corresponding effective couplings are part of set of 
processes to compose calculable observables.
Also, available calculations for the  electromagnetic  EH effective action 
neglect coupling(s) to fermions.
 In this work  
the large quark mass expansion of the quark determinant, that 
had been envisaged in Ref.
\cite{Novikov1984}
will be considered
and background quark sources will be introduced.
The general idea of this  method goes back to the Schwinger
operator method \cite{Schwinger1951}.
One may expect it  to be suitable  not only to heavy (charm and beauty)
 quark contributions but also to weak fields as compared to  low-energy quark constituent 
 masses.
Constituent quark masses are usually large since
they are obtained 
 by  mechanisms  of 
mass generation such as Dynamical Chiral Symmetry (DChSB)
and specific conditions for the validity of the expansion will be provided.
 As a natural consequence, the Abelian limit will be recovered
to provide the electromagnetic EH effective action in the presence of fermion sources
within a large fermion mass expansion.

%The EH-type effective action for  zero quark mass limit, or strong field, has been 
%explicitly calculated in \cite{Ozaki2014}
%expecting to correspond to a glasma phase that may occur in relativistic heavy.
% It has also been derived  in the presence  of  constant electromagnetic 
%fields, or magnetized,  massless QCD \cite{Galilo2011,EH-strongB,Ozaki2015a}.

In the following,  some  more specific  motivations for the present work are enumerated,
with, in addition to that,
 some problems  that can receive new insight or  benefit from our
calculations and results.
(1)
The imaginary part of the
electromagnetic EH, arising from a strong electric field,
 has been associated to instabilities that may drive pair creation \cite{Schwinger1951}.
The effect of explicit  coupling to fermions has never been analyzed and it may shed light in the description of such phenomenum.
Similarly it is interesting to understand whether QCD-EH effective action 
with explicit quark sources present any phenomena associated to these or other different
 instabilities.
(2)
Chromo-electric and chromo-magnetic properties of the  QCD ground state (vacuum) 
have been investigated in the recent years.
Results,  from both analytical and lattice QCD calculations,
 indicate that the Yang-Mills vacuum is rather chromo-magnetic and the c-electric field destabilizes it \cite{Savvidy1977,Savvidy2020,Bali2014}.
We expect to provide some insight into this subject by providing an estimation 
for the effect of the quark-antiquark condensate in such properties
what has received some attention in \cite{rafelskietal}.
(3) 
At the classical level, Lagrangian mass terms break scale symmetry,
being    quantum effects responsible for  its anomalous breaking
even in the 
case of massless fields classically \cite{Weinberg1995}.
It must be associated to a mass-generation mechanism.
The contribution of trace anomaly to the electron mass
was calculated in \cite{QED-traceanomaly}
and it has been shown to be  needed for the 
understanding   of hadron masses \cite{nucleonmass}.
So, the inclusion of quark sources in the EH-type effective action
can provide hints of how trace anomaly effectively yields such contributions
from a dynamical point of view.
(4) Related  to this last issue, and another one below,
given the intrincate low energy QCD structure,
one is usually lead to develop low energy models, that in part have evolved to 
the idea of EFT.
Despite the non-linear character of effective actions of the type of the 
 EH- action,
its expansion in polynomials of the fields may in some cases
 offer a more transparent way to interpret  the physical content of the involved couplings
and their contribution to physical processes.
This is the case of light scattering.
All different  possible couplings 
between the relevant effective degrees of freedom in terms of the
fields,  allowed by symmetry arguments,
 should emerge 
\cite{Weinberg1995}. 
Besides that,  models are continuously useful to understand hadron observables
which sometimes miss a more direct relation to the fundamental theory.
So, by deriving a low energy EFT, 
it becomes more direct (or natural) the 
use of different (effective or phenomenological)
 couplings and their corresponding roots  in more fundamental degrees of freedom
from quarks and gluons.
When describing quark and gluon dynamics (or photon- charged fermion),
gauge field coupling to matter may involve (weaker) higher-order couplings
 besides the minimal   and the magnetic moment couplings. 
As well understood in QED,
different electromagnetic field configurations can induce
 different types of fermion interactions.
In QCD, quarks and gluons should have enormously more possible interaction structures.
Accordingly,
chromo-electric and chromo-magnetic properties of the ground state (vacuum), and the
corresponding gauge field interactions with matter, are much more difficult to be
investigated and to be assessed
and we expect to provide some insight into this subject.
(5) 
The resulting EFT 
can provide a framework for the calculation of  hadron structure 
as it has been already done in some works without quark sources in
 \cite{Araki-etal,Shuryak-1980,Franz-etal}. 
We do not provide, however,  estimations of possible contributions
 for hadron structure in this  work. 
In some cases,
an EFT is postulated by heuristic arguments
and latter derived from more fundamental degrees of freedom
as, for example, the large Nc Weinberg EFT for quarks and pions \cite{LargeNcWeinberg,EPJA-2016}.
In the present work, background quark currents are introduced by    taking them as 
 sources in equations of motion.
By background quark field or current, one must understand 
 fermion fields (and corresponding currents)
that 
  obey some 
Dirac equation, in the present case with the coupling(s) to the Yang Mills fields.

In this work, the one-loop  quark determinant,
for massive quarks, with  background gluon fields
 is considered in a large quark mass expansion
 by considering quark currents as sources.
This gauge invariant large quark mass expansion
% have been  presented in \cite{Novikov1984},
 will be calculated 
up to operators of dimension $M^8$ (mass)
in terms of  
quantities as $G^{\mu\nu}$,
$J_\mu$ and ${\cal D}^\mu J^\nu$.
The contributions of the gluon and ghost determinants will be left outside the scope.
As a result, besides the effective gauge invariant 
gluon self interactions
found in other works of the field, some effective 
higher order quark-gluon effective Lagrangian couplings
are found.
These low energy effective couplings
 may contribute for identifying effects and processes that can take place
in the complicated low-intermediary energies
quark and gluon  dynamics.
The work is organized as follows.
In the next section, the quark determinant is  expanded
by  writing it as a sum of a scalar-field type determinant and an intrinsically fermionic part \cite{Novikov1984}.
The Abelian limit is explicitly exhibited in Section 
\eqref{sec:abelian}.
Some properties and observables are  calculated in Section \eqref{sec:properties}.
Given the presence of imaginary parts from the coupling to quark currents,
  the corresponding conditions for their zero imaginary parts are analyzed separately for each
quantity or  observable.
The Abelian limits  of these observables
are also exhibited.
Some identities used along the derivation of the effective
action are shown in Appendices with 
the corresponding corrections to the classical (dynamical) Yang-Mills equations.
In the last section (\ref{sec:summary})
a summary is presented.

\section{ The quark determinant in a gluon background: large quark mass  or weak gluon field expansion }
\label{sec:quarkdet}

The QCD path integral can be written as:
\begin{eqnarray}
Z [\eta,\bar{\eta},J_\mu, \xi]
= \int D [\psi,\bar{\psi}] D [A_\mu]
e^{ i \int d^4 x
\left( {\cal L}_{QCD} + 
L_{g.f.g.} \right)},
\end{eqnarray}
where 
$\mathcal{L}_{g.f.g.}$ contains the 
sources ($\eta,\bar{\eta},J_\mu, \xi$), ghost terms, and the gauge fixing that will not be investigated in the present work.
The Lagrangian density 
is given by:
\begin{eqnarray}
\cal{L}_{QCD} &=&
\Bar{\psi}\left( \slashed{\Pi} - m \right) \psi 
- \frac{1}{2} \Tr_C{ G^{\mu\nu} G_{\mu\nu}},
\end{eqnarray}
where $\slashed{\Pi}  =
i \slashed{\partial}  + g \slashed{A}$
and the gluon stress tensor
$G^{\mu\nu} =
\partial^\mu A^\nu
- \partial^\nu A^\mu
+ i g [ A^\mu, A^\nu]$
for $A^\mu = A^\mu_a \cdot 
\lambda_a$,
$\lambda_a$ being   Gell Mann color matrices and
$a,b,c,..$  color indices.

The quark part of the QCD action can be integrated out of exactly.
It yields the following effective action:
\begin{align} \label{EffAction1}
    i\Gamma_{\mbox{\tiny q}} 
=\Tr \ln{ \left( \frac{\slashed{\Pi}-m}{\slashed{p}-m} \right)},
\end{align}
that has been normalized with respect to the free theory, 
where the momentum operator 
was written as $\slashed{\Pi} =
\slashed{p} + g \slashed{A}$,
and the trace $\Tr$ is taken in all internal and spacetime indices.
This expression naturally diverges, 
and a Pauli-Villars regulator will be introduced later to make it possible to renormalize the resulting effective action.

To proceed with the calculation
in a more clear way, 
the derivative of this effective action with respect to the quark mass will be taken. 
The action is then calculated through:
\begin{align}
\Gamma_{\mbox{\tiny q}} = i m 
    \; \Tr \;\lim_{\Lambda \rightarrow \infty} \int_\Lambda^m  \left( \frac{1}{\Pi^2 - m'^2 } - \frac{1}{p^2 - m'^2} \right) dm',
\end{align}
where $\Lambda$ reduces to the 
Pauli-Villars regulator.
The $\sigma_{\mu \nu}$ matrices are defined through the Dirac gamma matrices as:
\begin{align}
    \sigma_{\mu \nu}
    =
    \frac{1}{2} [\gamma_\mu, \gamma_\nu].
\end{align}

Equation \eqref{EffAction1} has been calculated exactly for the background constant field case in \cite{Ozaki2015a}. Here we employ the expansion introduced by Novikov et al \cite{Novikov1984} and calculate the corrections up to operators of [mass]$^8$. 
To do so, an expansion for $G_{\mu \nu} G^{\mu\nu}/m^4 << 1$ is made, which is why we call it the large mass or weak field limit. The first part of this expansion  can 
be written as:
\begin{align}\nonumber
    \Gamma_{\mbox{\tiny q}} = i m \lim_{\Lambda \rightarrow \infty} \int_\Lambda^m \Tr &\left( D - D_0 - D^2 \frac{ig }{2} \sigma_{\mu \nu} G^{\mu \nu} + D^2 \frac{ig }{2} \sigma_{\mu \nu} G^{\mu \nu} D \frac{ig }{2} \sigma_{\mu \nu} G^{\mu \nu} \right. \\\nonumber
    & - D^2 \frac{ig }{2} \sigma_{\mu \nu} G^{\mu \nu} D \frac{ig }{2} \sigma_{\mu \nu} G^{\mu \nu} D \frac{ig }{2} \sigma_{\mu \nu} G^{\mu \nu}\\\label{EffAction2}
    & + \left. D^2 \frac{ig }{2} \sigma_{\mu \nu} G^{\mu \nu} D \frac{ig }{2} \sigma_{\mu \nu} G^{\mu \nu} D \frac{ig }{2} \sigma_{\mu \nu} G^{\mu \nu} D \frac{ig }{2} \sigma_{\mu \nu} G^{\mu \nu} + \mathcal{O}(\mbox{[mass]}^8) \right) dm',
\end{align}
where $D=1/(\Pi^2-m'^2)$ and $D_0=1/(p^2-m'^2)$).

Note that the field $G_{\mu \nu}(x)$ 
does not commute with operators of $\Pi_\mu$, in particular those 
in $D$, so 
that they can 
be isolated into local operators of $x$ with coefficients containing the momentum integrations  $p$. This is done by using the commutators found in Appendix A for which we found some small differences with respect to the equations presented in \cite{Novikov1984}. Equation \eqref{EffAction2} is the starting point for the 
derivative expansion. 
The effective action can be more compactly written as:
\begin{eqnarray}\label{DecompositionScalarAndFermion}
\frac{d \Gamma_q}{d m^2} =   \frac{i}{2} 
 \Tr\; 
\left[ 
(D - D_0 )
+
\sum_{n=1}^\infty 
\; (-1)^{n} D \left( \frac{ig}{2} \sigma_{\alpha \beta} G^{\alpha \beta}  D \;\right)^n \right]
\equiv 
I_{s} + I_{f}
=  \sum_n I_{s,n}
+
\sum_n I_{f,n},
\end{eqnarray}
where $I_s$ does not have Fermionic structures, while $I_f$ contains all the terms from the Dirac parts of Eq. \eqref{DecompositionScalarAndFermion}.
The first term 
is the same as a   bosonic scalar particle interacting with the gluon field, corresponding therefore to 
the scalar boson determinant.
Therefore, it will be called the  \textit{scalar QCD}.
This term will be used for the action with the multiplication of $-2$ due to the inverse squared root that appears in the 
scalar field determinant.
All the resulting terms will be
considered in the local limit.

The expansion will be done up to operators of order [mass]$^8$,
corresponding to terms such as
$G^4$ or $G^{\mu\nu} J^\mu J^\nu$,
where $J^\mu = \lambda_a \cdot 
\bar{\psi} \gamma^\mu \lambda^a \psi$ is the quark current.
Besides that, several terms will depend on the
covariant derivative ${\cal D}$,
such that the expansion and the final results will always be written
in terms of gauge invariant quantities:
\begin{eqnarray}\label{OPE}
\Gamma_{\mbox{\tiny q}}
= \Tr_{\mbox{\tiny C,F,L}} \sum_{ijk} c^{ijk}
(G_{\mu\nu})^i ({\cal D}_\alpha)^j
(J_\beta)^k,
\end{eqnarray}
where $c^{ijk}$ are the coefficients of the expansion, with dimensions of [mass]$^{-(2i+j+3k)}$. 
This effective action generalizes the above one $\tilde{\Gamma}_q$ since it includes
fermion currents.
We stress that the derivative appearing in the expansion above is the covariant derivative in the adjoint rep. This 
leads to a crucial difference from the covariant derivative appearing in the original action $\Pi_\mu/i$, as we can see, for instance, 
when acting on a 
generic functional built out of $G_{\mu\nu}$ such as 
$F(G)$:
\begin{eqnarray}
\label{DDcFG}
    [\mathcal{D}_\mu, \mathcal{D}_\nu] F(G) = &   -ig 
    [ G_{\nu \mu} , F(G) ] .
\end{eqnarray}

%a general scalar 
%field $\phi$:
%\begin{align}
%    [D_\mu, D_\nu] 
%\phi = &  ig 
%G_{\mu \nu} \phi;\\
%  [\mathcal{D}_\mu, 
%\mathcal{D}_\nu] = & %-ig [G_{\mu \nu}, %\phi].
%\end{align}

We have two different conditions on the index in the possible terms of the operator expansion \eqref{OPE}: one to ensure the mass order and the other from the requirement that each operator forms a Lorentz scalar. These two conditions are respectively given by:
$$
2i + j + 3 k \leq 8, 
\;\;\;\;\;\;
2 i + j + k = \mbox{even}.
$$
A list of all the possible terms according to these criteria,
 and taking into account the gauge invariance,
  is shown in Appendix  \ref{app:A}.
Several resulting terms can be written in different ways, leading to a partial ambiguity in the sector of quark current interactions.
All these terms show up in the resulting expanded effective action below.

\subsection{Expansion of the scalar part}

The $I_s$ part of Eq. \eqref{DecompositionScalarAndFermion} can be developed by using the translational invariance trick. Since the expressions for the derivative of the action are convergent, we can translate the momentum by $p_\mu \rightarrow p_\mu - q_\mu$, and, by expanding up to the second order in $q$, the following  identity is obtained:
\begin{align}\nonumber
    \Tr \left( D - D_0 \right) =& \Tr \left( \frac{1}{(\Pi_\mu - q_\mu)^2 - m^2} - \frac{1}{(p_\mu - q_\mu)^2 - m^2} \right)
    \\
    \nonumber
    =& \Tr \left( D - D_0 \right) + \Tr \left( D 2 \Pi_\mu D - 2 D_0 p_\mu D_0\right)q^\mu
     + \frac{1}{2} \Tr \left( D 2 \Pi_\nu D 2 \Pi_\mu D + 2g_{\mu \nu} D^2 + D 2\Pi_\mu D 2\Pi_\nu D \right.
     \\ 
     &\left.-
      D_0 2 p D_0 2 p D_0 - 2g_{\mu \nu} D_0^2 + D_0 2p_\mu D_0 2p_\nu D_0 \right)q^\mu q^\nu + ...
\end{align}

Simple integration arguments can be used to conclude that each coefficient of powers of $q_\mu$ must identically vanish. In particular,  the coefficient of the second-order term will be 
multiplied  by $g^{\mu \nu}$, and we may rewrite this term using the commutators in Appendix:
\begin{align}\nonumber
    \Tr \left( D^3 \Pi^2 - D^2 - D_0^3 p^2 + D_0^2\right) =& - \Tr \left( D^2 [\Pi^2,\Pi_\mu] D \Pi^\mu \right)\\
    \Tr \left( D^3 - D_0^3 \right) =& - \frac{1}{m^2} \Tr \left( D^2 [\Pi^2,\Pi_\mu] D \Pi^\mu \right),
\end{align}
where we made use of the fact that $[D,F]=-D[\Pi^2,F]D$.

The equation above is precisely the second derivative of the scalar part in Eq. \eqref{DecompositionScalarAndFermion}. Here we pass all the $D$ operators to the right, and thus write:
\begin{align}\nonumber
    \frac{d^3 \Gamma_q}{d(m^2)^3} =& -\frac{i}{m^2} \Tr \left( D^4 [\Pi^2,\Pi_\mu] \Pi^\mu - D^5 [\Pi^2,\Pi_\mu] [\Pi^2,\Pi^\mu] + D^6 [\Pi^2,\Pi_\mu] [\Pi^2, [\Pi^2,\Pi^\mu]] \right.\\\nonumber
    &\left. - D^7 [\Pi^2,\Pi_\mu] [\Pi^2, [\Pi^2, [\Pi^2,\Pi^\mu]]] + D^8 [\Pi^2,\Pi_\mu] [\Pi^2, [\Pi^2, [\Pi^2, [\Pi^2,\Pi^\mu]]]] \right. \\\label{ScalarQCDS}
    &\left. - D^9 [\Pi^2,\Pi_\mu] [\Pi^2,[\Pi^2, [\Pi^2, [\Pi^2, [\Pi^2,\Pi^\mu]]]]] + \mathcal{O}([\mbox{mass}]^9) \right).
\end{align}

We listed all the terms that can contribute to [mass]$^8$. The appearance of the operators $\mathcal{D}_\mu$, $G_{\mu \nu}$ and $J_\mu$ comes from the commutators, but also remember that the $A_\mu$ field is still present in the scalar propagator $D = 1/ (\Pi^2-m^2)$. Therefore, we must supplement the commutators in Appendix \ref{Apendice_Comutadores} with the identities $\Tr \left( D^n F(\mathcal{D}, G, J) \right)$ up to [mass]$^4$, $\Tr \left( D^n \Pi^\alpha F_\alpha(\mathcal{D}, G, J) \right)$ up to [mass]$^5$, $\Tr \left( D^n \{ \Pi^\alpha, \Pi^\beta \} F_{\alpha \beta}(\mathcal{D}, G, J) \right)$ up to [mass]$^4$, $\Tr \left( D^n \Pi^\alpha \Pi^\beta \Pi^\mu F_{\alpha \beta \mu}(\mathcal{D}, G, J) \right)$ up to [mass]$^5$ and $\Tr \left( D^n \Pi^\alpha \Pi^\beta \Pi^\mu \Pi^\nu F_{\alpha \beta \mu \nu}(\mathcal{D}, G, J) \right)$ up to [mass]$^2$, which can be found in the Appendix \ref{Apendice_Identidades}. 

This case can be called \textit{scalar QCD} and must be addressed carefully. 
By that, we mean that the resulting corresponding terms for the 
action 
%$\Bar{\psi}(\slashed{\Pi} 
%- m) \psi$ 
can be considered to be the same as the resulting determinant from the Klein-Gordon action of a (real) scalar field $\phi$, \textit{i.e.} $\phi^\dagger (- \Pi^2 - m^2) \phi$.
The corresponding scalar determinant, with the same normalization chosen for the QCD case, is given by
\begin{align}
   \Gamma_{\mbox{\tiny sQCD}} = - \frac{1}{2} \Tr_{\mbox{\tiny C,F,p}} \ln \left( \frac{-\Pi^2-m^2}{-p^2-m^2} \right), 
\end{align}
where, now, we make explicit that there is no  trace 
in Dirac indices, \textit{i.e.} $\Tr_{\mbox{\tiny C,F,p}}$ 
for momentum, color and flavor indices.
Therefore, we can refer to it as scalar QCD provided that when adding to the strict fermionic part
Eq. \eqref{DecompositionScalarAndFermion} it is supplemented by the trace in Dirac indices with the above extra multiplicative factor $(-1/2)$, 
This action is not yet written in terms of fermion currents so that it is denoted by
$\tilde\Gamma_s$ instead of $\Gamma_s$.

\subsection{Expansion of the Fermionic part}

For the Fermionic $I_f$ in Eq. \eqref{DecompositionScalarAndFermion}, 
terms already contain structures with $G_{\mu \nu}$ field explicitly.
It can be seen that the term $I_{f,4}$ is already at the desired order in $G_{\mu\nu}$.
The terms in $I_f$ can immediately be written by taking the traces in Dirac indices:
\begin{align}\nonumber
    I_f = \Tr_{\mbox{\tiny C,F}} \int_{k} &\left( -ig^2 D^2 G^{\mu \nu} D G_{\mu \nu} - 2g^3 D^2 G_{\mu \nu} D G^{\nu}_{\;\; \alpha} D G^{\alpha \mu} \right.\\\label{I_F}
    &\left. + i D^5 \left( \frac{3}{2} G^4 + \frac{1}{2^2} [G_{\mu \nu}, G_{\alpha \beta}]^2 - \frac{1}{2} \{ G_{\mu \nu}, G^{\nu \alpha} \}^2 - \frac{3}{2} [G_{\mu \nu}, G^{\nu \alpha}]^2 \right) \right)
\end{align}

To develop the first two terms to [mass]$^8$,  
the structures in $G^{\mu\nu}$ can be isolated again by employing the commutators, and this yields the following equation:
\begin{align}
\nonumber
    I_f = \Tr_{\mbox{\tiny C,F}} \int_{k} &\left( -ig^2 D^3 \left\{ G^2 - D G^{\mu \nu} [\Pi^2, G_{\mu \nu}] - D^2[\Pi^2, G^{\mu \nu}][\Pi^2, G_{\mu \nu}] - D^3[\Pi^2, [\Pi^2, G^{\mu \nu}]][\Pi^2, G_{\mu \nu}]  \right. \right. \\\nonumber
    &\left.  - D^4[\Pi^2, [\Pi^2, [\Pi^2, G^{\mu \nu}]]][\Pi^2, G_{\mu \nu}] \right\}  - 2g^3 D^4 \left\{ G_{\mu \nu} G^{\nu}_{\;\; \alpha} G^{\alpha \mu} + D \left( [\Pi^2, G_{\mu \nu}]G^{\nu}_{\;\; \alpha} G^{\alpha \mu} \right. \right. \\\nonumber
    &\left. \left. - G_{\mu \nu} G^{\nu}_{\;\; \alpha} [\Pi^2,G^{\alpha \mu}] \right) - D^2 \left( [\Pi^2,G_{\mu \nu} G^{\nu}_{\;\;\alpha}] [\Pi^2,G^{\alpha \mu}] + [\Pi^2, G_{\mu \nu},[\Pi^2, G^{\nu}_{\;\; \alpha} G^{\alpha \mu}]] \right) \right.\\ \label{fermionpartS}
%\label{I_f 2}
    &\left. \left. - D^2 [\Pi^2, G_{\mu \nu}] G^{\nu}_{\;\; \alpha} [\Pi^2,G^{\alpha \mu}] \right\} + i D^5 \left( \frac{3}{2} G^4 + \frac{1}{2^2} [G_{\mu \nu}, G_{\alpha \beta}]^2 - \frac{1}{2} \{ G_{\mu \nu}, G^{\nu \alpha} \}^2 - \frac{3}{2} [G_{\mu \nu}, G^{\nu \alpha}]^2 \right) \right).
\end{align}

Here we stress that the equation above can be written in different ways, depending on if we pass the $D$ operator in Eq. \eqref{I_F} to the left, to the right, or mix these ways in the calculation. Note that all but the first term in Eq. \eqref{fermionpartS} are UV finite when we perform the momentum integrals, so only the first term will renormalize the charge.
%Later,  a contribution of 
%the gluon determinant can 
%also be considered.

\subsection{ Complete fermionic action: EFT }

Next, we present the resulting local limit of the
effective action after resolving all the commutators and by adding the terms from the fermionic part 
 -Eq. \eqref{fermionpartS} - to the 
 scalar QCD part - Eq. \eqref{ScalarQCDS}.
 For that, 
 the identities
 shown in 
\eqref{Apendice_Identidades} are employed
and an example of explicit calculation is presented in 
Appendix \eqref{app:coefficients}.
Let us denote
the resulting dimensionless coefficients
(coupling constants) after integrations in momentum
of the fermionic part by 
$\zeta_i$ ($i=1,..10$) and the ones from
the sQCD by $g_i$.
As noted above, they are related to the QCD couplings $\lambda_i$ (that must not be confused with 
Gell Mann matrices) via:
\begin{align}
\label{sumcouplings}
    \lambda_i = -2 g_i + \zeta_i,
\end{align}
 The final result, in the local limit,
 is 
given by:
\begin{eqnarray}
\label{Seffresult}
\nonumber
    \Gamma_{\mbox{\tiny q}} &=& \int d^4x \left\{-\frac{1}{4} \left( 1 + \frac{2 N_{\mbox{\tiny F}}g^2}{48 \pi^2} \ln{\frac{\Lambda^2}{m^2}} \right) (G_a^{\mu \nu})^2 + \Bar{\psi} \left( \slashed{\Pi}-m \right) \psi 
    \right.
    \\
    \nonumber
    &+& \frac{\lambda_1}{m^2} f_{abc} G_a^{\mu \nu} G_{\nu \rho b} G^{\rho}_{\mbox{ } \mu c} +  \frac{\lambda_3}{m^4} k_{abcd} G_{\mu \nu a} G^{\mu \nu}_b G_{\alpha \beta c} G^{\alpha \beta}_d + \frac{\lambda_4}{m^4} f_{abc} f_{ade} G_{\mu \alpha b} G^{\alpha \nu}_c G^{\mu \beta}_{d} G_{\beta \nu e} 
    \\\nonumber
    &+&  \frac{\lambda_5}{m^4} f_{abc} f_{ade} G_{\mu \nu b} G_{\alpha \beta c} G^{\mu \nu}_{d} G^{\alpha \beta}_e + \frac{\lambda_6}{m^4} k_{bcde} G_{\mu \alpha b} G^{\alpha \nu}_c G^{\mu \beta}_d G_{\beta \nu e}  
    \\ 
    \nonumber
    &+&
      \frac{\lambda_7}{m^4} f_{abc} (\mathcal{D}^\mu J_\nu)_a G_{\mu \alpha b} G^{\alpha \nu}_c + \frac{\lambda_8}{m^4} d_{abc} \left( \mathcal{D}^\mu J_\nu \right)_a G_{\mu \alpha b} G^{\alpha \nu}_c
    \\ 
    &+& \left. 
     \frac{\lambda_2}{m^2} J^\mu_a J_{a \mu}
     + 
 \frac{\lambda_9}{m^4}f_{abc} G_{\mu \nu a} J^\mu_b J^\nu_c    
    + \frac{\lambda_{10}}{m^4} \left(\mathcal{D}_\alpha J_\mu \right)_a \left(\mathcal{D}^\alpha J^\mu\right)_a \right\},
\end{eqnarray}
where
$f_{abc}$ ($d_{abc}$) are the antisymmetric (symmetric) structure constants for SU(3) \cite{SU3} and
the following tensor was defined:
\begin{align}
    k_{abcd} = \frac{\delta_{ab} \delta_{cd}}{3} + \frac{d_{abe}d_{cde}}{2}.
\end{align}
Note there is no need of redefining further the effective action
since quark currents showed up due to the equation of motion Eq. \eqref{EOM}.

The resulting structures from the expansion of the 
scalar and purely fermionic parts are the same, and they add to each other to form the above expression.
The only exception is the correction to the
effective quark interaction
with coupling constant
$\lambda_2$ that arises 
exclusively from sQCD part.
This term is the leading interaction term in the 
large quark mass expansion ($\sim \lambda/m^2$) although its coupling constants is not larger than the others.
Since we are exploiting the local limit of the effective action, the term $\lambda_2$ can be seen as a 
one-loop correction to effective 
contact interaction due to vacuum polarization that should add to different other contributions \cite{NambuJonaLasinio,Weise}.
Different types of dynamical couplings show up:
momentum dependent fourth  order gluon interactions ($G^4$)
- $\lambda_3, \lambda_4, \lambda_5, \lambda_6$ - (and corresponding to higher order gluon interactions
${A^\mu}^n$ for $ 4 \leq n \leq 8$).
Two gluon fusion into  one or one gluon splitting into two gluons in the coupling $\lambda_1$, that also carries higher order terms with  $A_\mu^m$
with $3 \leq m \leq 6$.
The quark-gluon interactions contain several types of quark current-gluons couplings 
($\lambda_7, \lambda_8$) and also quark current interactions with at least one gluon external line ($\lambda_9, \lambda_{11}$).
 For constant chromo-electric or chromo-magnetic fields, these terms also represent corrections to effective four-quark contact interactions.
Concerning the pure gluonic terms,
some of the terms above are manifestation of the non-Abelian character of the gluon dynamics, $\lambda_1, \lambda_4$, and $\lambda_5$.
Other terms survive in the Abelian limit, as shown below.
The terms $\lambda_3$, $\lambda_4$ and $\lambda_5$, $\lambda_6$
can be written in different ways, being that they may add differently in each way and this is due to their non-Abelian content.
Although the exclusive gluon field contributions for this resulting Euler-Heisenberg-type QCD-effective action
do not have an imaginary part,
the gauge field interaction with fermions has one:
the $\lambda_8$ term.
The regime of pair-creation can be then analyzed.
Furthermore, seen in a different channel, these terms may describe the emission and absorption of gauge field configurations seen below.

 Numerical values for the dimensionless couplings $g_i$, 
$\zeta_i$ and $\lambda_i$, of Eq. \eqref{Seffresult}, can be found in
% first three columns of 
Table \eqref{tab:couplingsSeff}.
  As it is pointed  out in Ref.  \cite{Novikov1984}, there are several conflicting results in the literature for the 
resulting parameters of  the gluon sector (i=1,3,4,5,6).
For the sake of comparison, 
 the values of the effective coupling constants not equal to those 
presented in the Table, 
 obtained
in Ref. \cite{Novikov1984}  
 are the following:
\begin{eqnarray}
\zeta_1 &=& +\frac{3g^3}{10520 \pi^2},
\nonumber
\\
g_4 &=& +\frac{g^4}{64512 \pi^2}
,
\;\;\;\;
\zeta_4 = - \frac{51 g^4}{71680 \pi^2},
 \;\;\;\;
\lambda_4 = + \frac{29 g^4}{40320 \pi^2},
\nonumber
\\
g_5 &=& +\frac{g^4}{645120 \pi^2}, \;\;\;\;
\zeta_5 = + \frac{289g^4}{ 1290240 \pi^2},\;\;\;\;
 \lambda_5  =- \frac{g^4}{2240 \pi^2},
\end{eqnarray}
 For some terms, the two contributions have opposite signs, although one cannot forget that
 the relative sign of each contribution  (Dirac part and sQCD part) are opposite - see Eq. \eqref{sumcouplings}.
 Therefore, it is interesting to see how these effective interactions in full
 QCD are different from the ones in scalar-QCD.
 The coupling constants do not show to be numerically convergent at this order of the expansion, so the convergence of the series can be assured by requiring small quantities of the type:
\begin{eqnarray} \label{convergence1}
\frac{G^{\mu \nu} \cdot G_{\mu \nu} }{m^4},
\;\;\; \frac{{\cal D}^\nu J^\mu}{m^4},
\;\;\; \frac{J^\mu J^\nu}{m^6}
\;\; < \; 1.
\end{eqnarray}

So, let us consider typical values for 
heavy quark  (HQ) masses ($m \sim 1.7$ GeV) and
for light constituent quark (LCM) $m \sim 0.35$ GeV.
Besides that, the quark currents can be exchanged by
scalar densities by a Fierz transformation and then to generate
the scalar quark-antiquark condensate in the vacuum, 
nearly  as $J^\mu J_\mu \sim < \bar{\psi} \psi >^2$.
The following conditions for the validity of the resulting expansion can be considered
\begin{eqnarray}  \label{convergence2}
HQ && G^{\mu \nu}  \cdot G_{\mu \nu}  < 8.4 GeV^4,
\nonumber
\\
LCM &&  G^{\mu \nu} \cdot  G_{\mu \nu}  < 0.015 GeV^4, \;\;\;\;\; \mbox{and}
\;\;\;\;
\frac{< \bar{\psi} \psi >}{ m^3} \sim 0.36 < 1.
\end{eqnarray}
Whereas the limit for HQ  is  well compatible with the expansion in a large range of values,
the LCM case is more subtle.
Usual values for the constituent quark mass may be in the desired range of values 
for small values of $G^{\mu\nu} G_{\mu\nu}$,
for example with  typical values of the vacuum gluon condensate 
$ < G^{\mu \nu} G_{\mu \nu}> \sim  2 \Lambda_{QCD}^4$ \cite{Savvidy2020}.
The comparison of typical values of the  constituent quark mass
is also reasonable with respect to the scalar quark condensates,
 $< \bar{\psi} \psi > \sim (-0.25)^3$ GeV$^3$
\cite{condensates}.

{\color{red}

\begin{table}[H]
\centering
\caption{ Resulting dimensionless coupling constants for the scalar QCD ($g_i$), the fermionic part ($\zeta_i$)
%, Eq. (\ref{fermionpart}),
and the sum 
$\lambda_i = g_i + (-2)\zeta_i$. 
%The last three columns are the results from \cite{Novikov1984}.
}
\resizebox{.5\columnwidth}{!}{
\begin{tabular}{l||l|l|l}
i &  \; $g_i$ \; & \; $\zeta_i$
\; & \; $\lambda_i$ \; 
%& \; $g_i$ (\cite{Novikov1984}) \;& \; $\zeta_i$ \cite{Novikov1984} \;& 
%\; $\lambda_i$ \cite{Novikov1984} \; 
\\ 
\hline
i = 1 
    & $+\frac{g^3}{5760 \pi^2}$ 
    & $+\frac{71g^3}{10944 \pi^2}$
%$\zeta_1$ 
    & $+\frac{7g^3}{1440\pi^2}$
 %   & $+ \frac{g^3}{5760 \pi^2}$
 %   & $+\frac{3g^3}{10520 \pi^2}$
 %   & $- \frac{g^3}{2880 \pi^2}$
\\
& & & 
%& & &
\\
i = 2 
    & $+\frac{13g^2}{34560 \pi^2}$ 
    & $0$
%$\zeta_2$ 
    & $-\frac{13g^2}{17280\pi^2}$
%    & $-$
%    & $-$
%    & $-$
\\
& & & 
%& & &
\\
i = 3 
    & $+ \frac{g^4}{18432 \pi^2}$ 
    & $-\frac{g^4}{3072 \pi^2}$ 
%$\zeta_3$
    & $-\frac{g^4}{2304\pi^2}$
%    & $+\frac{g^4}{18432 \pi^2}$
%    & $- \frac{g^4}{3072 \pi^2}$
%    & $- \frac{g^4}{2304 \pi^2}$
\\
& & & 
%& & &
\\
i = 4
    & $-\frac{47g^4}{161280\pi^2}$ 
    & $- \frac{9g^4}{2560 \pi^2}$ 
%$\zeta_4$ 
    & $-\frac{473g^4}{161280\pi^2}$ 
 %   & $+\frac{g^4}{64512 \pi^2}$
 %   & $- \frac{51 g^4}{71680 \pi^2}$
 %   & $+ \frac{29 g^4}{40320 \pi^2}$
\\
& & &
% & & &
 \\
i = 5
    & $+\frac{g^4}{23040 \pi^2}$ 
    & $+ \frac{43 g^4}{80640 \pi^2}$ 
%$\zeta_5$
    & $+\frac{g^4}{2240\pi^2}$ 
%    & $+\frac{g^4}{645120 \pi^2}$
%    & $+ \frac{289g^4}{ 1290240 \pi^2}$
%    & $- \frac{g^4}{2240 \pi^2}$
\\
& & & 
%& & &
 \\
i = 6 
    & $+ \frac{g^4}{23040}$ 
    & $+\frac{g^4}{768 \pi^2}$ 
%$\zeta_6$ 
    & $+\frac{7g^4}{5760\pi^2}$
 %   & $+\frac{g^4}{23040 \pi^2}$
 %   & $+\frac{g^4}{768 \pi^2}$
 %   & $+ \frac{7g^4}{5760\pi^2}$
\\
& & &
% & & &
 \\
i = 7 
    & $- \frac{79g^3}{6912\pi^2}$ 
    & $- \frac{2863 g^3}{92160 \pi^2}$ 
%$\zeta_7$ 
    & $-\frac{2269g^3}{276480\pi^2}$
%    & $-$
%    & $-$
%    & $-$
\\
& & & 
%& & & 
  \\
i = 8
    & $+\frac{29ig^3}{35840\pi^2}$ 
    & $-\frac{41ig^3}{10240 \pi^2}$ 
%$\zeta_8$
    & $-\frac{403ig^3}{71680\pi^2}$  
 %   & $-$
 %   & $-$
 %   & $-$
\\
& & &
% & & &
  \\
i = 9 
    & $+\frac{97g^3}{3870720 \pi^2}$ 
    & $- \frac{49g^3}{92160 \pi^2}$ 
%$\zeta_9$
    &  $-\frac{563g^3}{967680\pi^2}$
%    & $-$
%    & $-$
%    & $-$
\\
& & & 
%& & &
  \\
i = 10 
    & $+\frac{1609g^2}{967680\pi^2}$ 
    & $+ \frac{g^2}{1920 \pi^2}$ 
%$\zeta_{10}$ 
    & $-\frac{1357g^2}{483840\pi^2}$ 
%    & $-$
%    & $-$
%    & $-$
%\\
%& & & 
\end{tabular}
}
\label{tab:couplingsSeff}
\end{table}

}

In Eq. \eqref{Seffresult} there is one single UV divergent structure ($G^2$) - for instance, the first term. It is eliminated by  a subtraction of the kinetic gluon term, and it  
brings the need to renormalize the (running) coupling constant. With the Pauli-Villars subtraction \eqref{EffAction1}, by generalizing to $N_{\mbox{\tiny F}}$ quarks, the renormalized coupling constant becomes, respectively for scalar QCD ($g_R^2$) and QCD ($\lambda_R^2$):
\begin{align}
g_R^2 =& \frac{g^2}{
1 + \frac{2 N_{\mbox{\tiny F}} g^2}{96 \pi^2} \ln \left(
\frac{\mu^2}{m^2}\right)
}; \\
\lambda_{R}^2 =& \frac{\lambda^2}{
1 + \frac{2 N_{\mbox{\tiny F}} \lambda^2}{48 \pi^2} \ln \left(
\frac{\mu^2}{m^2}\right)
}.    
\end{align}
The scalar QCD part contributes to half of the complete one-loop fermionic divergent term.
The second result is the usual fermionic part of the complete QCD one-loop running coupling constant 
\cite{Peskin2018}, which does not exhibit the 
complete behavior because of the absence of the gluon 
contribution. 

% \begin{table}[H]
% \centering
% \label{couplingsg}
% \caption{  Results from \cite{Novikov1984} for the dimensionless coupling constants for the scalar QCD ($g_i$), the fermionic part $\zeta_i$, Eq. (\ref{fermionpart}),
% and the sum 
% $\lambda_i = g_i + (-2)\zeta_i$.
% }
% \resizebox{.3\columnwidth}{!}{\begin{tabular}{l||l|l|l}
% i &  \; $g_i$ \; & \; $\zeta_i$
% \; & \; $\lambda_i$
% \\ 
% \hline
% i = 1 & $+\frac{g^3}{5760 \pi^2}$ 
%       & $ + \frac{3 g^3}{10520 \pi^2} $
% %$\zeta_1$ 
% & 
% $-\frac{g^3}{2880\pi^2}$  
% \\
% & & & 
% \\
% i = 3 & $+ \frac{g^4}{18432 \pi^2}$ 
%       & $-\frac{g^4}{3072}$ 
% %$\zeta_3$
% & $-\frac{g^4}{2304\pi^2}$
% \\
% & & & 
% \\
% i = 4
% & $+\frac{g^4}{64512\pi^2}$ 
% & $ - \frac{51}{71680 \pi^2} $ 
% %$\zeta_4$ 
% & $\frac{29g^4}{40320\pi^2}$  
% \\
% & & & 
%  \\
% i = 5
% & $ + \frac{g^4}{645120\pi^2} $ 
% & $ + \frac{289 g^4}{1290240 \pi^2} $ 
% %$\zeta_5$
% & $ - \frac{g^4}{2240 \pi^2} $ 
% \\
% & & & 
%  \\
% i = 6 
% & $ + \frac{g^4}{23040 \pi^2} $ 
% & $ + \frac{g^4}{768 \pi^2} $ 
% %$\zeta_6$ 
% & 
% $ + \frac{7g^4}{5760\pi^2} $  
% \end{tabular}
% }
% \end{table}

\subsection{ The
  Abelian limit }
\label{sec:abelian}

The above effective action reproduces Euler-Heisenberg's results for QED \cite{EulerHeisenberg1935, Schwinger1951} up to the fourth order.
For colorless fields with one fermionic component $N_F=1$,
 we make the following identifications:
\begin{align}
    \Tr_{\mbox{\tiny C}} \left( G^4  \right) =& \frac{k_{abcd}}{4} G_{\mu \nu a} G^{\mu \nu}_b G_{\alpha \beta c} G^{\alpha \beta}_d \rightarrow (F^2)^2;\\
    \Tr_{\mbox{\tiny C}} \left( \{ G_{\mu \nu}, G^{\nu \alpha} \}^2 \right) =& k_{abcd} G_{\mu \nu a} G^{\nu \alpha}_b G^{\mu \nu}_c G_{\nu \alpha d} \rightarrow 4 (F_{\mu \nu} F^{\nu \alpha})^2 ;\\
    \Tr_{\mbox{\tiny C}} \left( J_\mu J^\mu \right) =& \frac{1}{2} J_{\mu a} J^{\mu}_a \rightarrow j^2;\\
    \Tr_{\mbox{\tiny C}} \left( (\mathcal{D}^\mu J_{\nu}) \{ G_{\mu \alpha}, G^{\alpha \nu} \} \right) =& \frac{1}{2} d_{abc} (\mathcal{D}^\mu J_{\nu})_a G_{\mu \alpha b} G^{\alpha \nu}_c \rightarrow 2(\partial^\mu j_\nu) F_{\mu \alpha} F^{\alpha \nu};\\
    \Tr_{\mbox{\tiny C}} \left( (\mathcal{D}_\mu J_\alpha) \mathcal{D}^{\mu} J^{\alpha} \right) =& \frac{1}{2} (\mathcal{D}_\mu J_\alpha)_a ( \mathcal{D}^{\mu} J^{\alpha} )_a \rightarrow (\partial_\mu j_\alpha)^2,
%\\
%   J^\mu_a \rightarrow& 0;\\
% N_F \rightarrow& 1,
\end{align}
where $F^{\mu \nu}$ is the electromagnetic strength tensor and $j_{\mu} = q \Bar{\psi} \gamma_\mu \psi$ the electromagnetic current. With these,
the action is written as:
\begin{align}\nonumber
S_{\mbox{\tiny eff}}^{\mbox{\tiny EH+j}} =& \int d^4x \left( -\frac{1}{4} \left( 1 + \frac{g^2}{24\pi^2} \ln{\frac{\Lambda^2}{m^2}} \right) F^2 +  \frac{4 \lambda_3}{m^4} (F^2)^2 + \frac{4 \lambda_6}{m^4} (F^{\mu \nu}F_{\nu \alpha})^2 \right.\\
&\left. + \; \frac{2\lambda_2}{m^2} j_\mu j^\mu + \frac{4\lambda_8}{m^4} (\partial_\mu j^\nu) F^{\mu \alpha} F_{\alpha \nu} + \frac{2 \lambda_{10}}{m^4} (\partial_\mu j_\nu)^2 \right).
\end{align}
This is an extended Euler Heisenberg 
effective action in the presence of 
fermion currents.
It contains
momentum independent and dependent fermion-fermion effective four-point interactions (respectively $\lambda_2$
and $\lambda_{10}$) and
a momentum-dependent fermion current interaction with two
photon insertions $\lambda_8$.
%Although these terms were obtained in the 
%local limit, their origin is the non-local 
%vacuum polarization.

By making the identifications of the strength tensor with the electric $\mathbf{E}$ and magnetic $\mathbf{H}$ fields \cite{Itzykson2006}, and the four-current with the charge density $\rho$ and vector current $\mathbf{j}$:
\begin{align}\nonumber
    F^{0i} =& E^i;
    \\\nonumber
    F^{ij} =& \epsilon^{ijk} H^k;
    \\\label{EHem}
    j^{0} =& \rho,
\end{align}
we obtain:
\begin{align} \label{EHsources}
\nonumber
S_{\mbox{\tiny eff}}^{\mbox{\tiny EH}} =& \int d^4x \left( -\frac{1}{4} \left( 1 + \frac{g^2}{24\pi^2} \ln{\frac{\Lambda^2}{m^2}} \right) \left( \mathbf{H}^2 - \mathbf{E}^2 \right) + \frac{2\lambda_2}{m^2} \left( \rho^2 - \mathbf{j}^2 \right) + \frac{4 \lambda_3}{m^4} \left( \mathbf{H}^2 - \mathbf{E}^2 \right)^2 \right.\\\nonumber
& + \frac{4 \lambda_6}{m^4} \left( 2(\mathbf{H}^2- \mathbf{E}^2)^2 
+ 4 (\mathbf{H} \cdot \mathbf{E})^2 \right) 
+
\frac{4\lambda_8}{m^4} \left( - (\mathbf{E}  (\mathbf{E} \cdot \nabla) + \mathbf{H} (\mathbf{H} \cdot \nabla))  \cdot \mathbf{j} + \frac{\partial \rho}{\partial t} \mathbf{E}^2 + \mathbf{H}^2 \nabla \cdot \mathbf{j} \right. \\
&\left.\left.  
+ \mathbf{H} \times \mathbf{E} \cdot \left( \frac{\partial \mathbf{j}}{\partial t} + \nabla \rho \right) \right) + \frac{2 \lambda_{10}}{m^4} 
\left(
(\frac{\partial \rho}{\partial t})^2 
- (\frac{\partial \mathbf{j}}{\partial t})^2
- (\nabla \rho)^2
+ (\partial_i j_k) \cdot
(\partial_i j_k)
\right) 
\right).
\end{align}
The usual Euler-Heisenberg type  terms composing the higher order electromagnetic effective dynamics
 $\lambda_3$ and $\lambda_6$ bring effects such as light-light interaction and birefringence. 
There are 
higher order 
terms with 
momentum independent and dependent effective fermion self-interactions,
$\lambda_2, \lambda_{10}$.
Besides that, there are several  electromagnetic field
interactions with fermions in $\lambda_8$, 
that is imaginary.
These contributions are  dynamical 
in the sense of 
fermion density and current 
spatially
non-uniform or time-dependent.
They may be
 associated with pair production.
However, by seeing in a different channel, 
 the 
last term of $\lambda_8$, 
has  an  structure of an electromagnetic wave either emitted or absorbed by
a time
and spatial variation of the charge density/current.
These imaginary terms disappear 
either in the absence or in the presence of electric fields,
being the general condition of real effective action given by:
\begin{eqnarray}
{\cal I}m (S_{eff}^{|EH}) = 0  &\;\;\;
\to &
0 =  - \left[ \mathbf{E}  (\mathbf{E} \cdot \nabla) + \mathbf{H} (\mathbf{H} \cdot \nabla)\right]  \cdot \mathbf{j} + \frac{\partial \rho}{\partial t} \mathbf{E}^2 + \mathbf{H}^2 \nabla \cdot \mathbf{j} 
+ \mathbf{H} \times \mathbf{E} \cdot \left( \frac{\partial \mathbf{j}}{\partial t} + \nabla \rho \right).
\end{eqnarray}
This condition can be guaranteed, for example, by requiring homogeneous and time independent fermion density and current. 
By choosing different 
conditions for 
$\mathbf{H}=0$ or $\mathbf{E}=0$,
it may also be guaranteed in particular sectors, such as:
\begin{eqnarray}
{\cal I}m (S_{eff}^{EH}) = 0  &\;\;\;
\to &
\;\;\;\;\;\;\;\;
- \mathbf{H} (\mathbf{H} \cdot \nabla)  \cdot \mathbf{j} 
 + \mathbf{H}^2 \nabla \cdot \mathbf{j} = 0,
 \;\;\;\;\;\;
\mathbf{E} = 0,
\nonumber
\\
{\cal I}m (S_{eff}^{|EH}) = 0  &\;\;\;
\to &
\;\;\;\;\;\;\;\;
- \mathbf{E} (\mathbf{E} \cdot \nabla)  \cdot \mathbf{j} 
+ \frac{\partial \rho}{\partial t}
\mathbf{E}^2 = 0,
\;\;\;\;\;\;\;\;\;\;\;\;\;\;\;
\;
\mathbf{B} = 0.
\end{eqnarray}
Non-homogeneous or time-varying charge or current densities also have particular contributions with $\lambda_{10}$ self-interaction terms.

In the absence of sources, 
$J_\mu= 0$,
 the weak field limit of the Euler-Heisenberg action 
 is recovered
\cite{EulerHeisenberg1935, Schwinger1951}:
\begin{align}
    S_{\mbox{\tiny eff}}^{\mbox{\tiny EH}} =& \int d^4x \left( -\frac{1}{4} \left( 1 + \frac{g^2}{24\pi^2} \ln{\frac{\Lambda^2}{m^2}} \right)(\mathbf{H}^2 - \mathbf{E}^2) + \frac{2\alpha^2}{45 m^4} \left( (\mathbf{H}^2 - \mathbf{E})^2 + 7 (\mathbf{H} \cdot \mathbf{E})^2 \right) \right).
\end{align}

\section{ Some properties of QCD-EH effective action}
\label{sec:properties}

In this section, some properties of Eq. \eqref{Seffresult} are shown.
For that, the
c-electric and c-magnetic fields are defined in the same way done in electromagnetism \eqref{EHem}
by:
\begin{eqnarray}
G^{0i}_a = E_a^i,
\;\;\;\;\;
G^{ij}_a = \epsilon^{ijk} H_a^k.
\end{eqnarray}

\subsection{
Chromo-electric and chromo-magnetic fields
}

The above resulting  effective action  can be rewritten as:
\begin{eqnarray}\nonumber
    \Gamma_{\mbox{\tiny q}} &=& \int d^4x \left\{-\frac{1}{2} \left( 1 + \frac{2 N_{\mbox{\tiny F}}g^2}{48 \pi^2} \ln{\frac{\Lambda^2}{m^2}} \right) \left( \mathbf{H}_a^2 - \mathbf{E}_a^2 \right) + \Bar{\psi} \left( \slashed{\Pi}-m \right) \psi 
    \right.
    \\
    \nonumber
    &+& \frac{\lambda_1}{m^2} f_{abc} \left( 3 \left( \mathbf{E}_a \times \mathbf{E}_b \right) \cdot \mathbf{H}_c - \left( \mathbf{H}_a \times \mathbf{H}_b \right) \cdot \mathbf{H}_c \right) +  \frac{4\lambda_3}{m^4} k_{abcd} \left( \mathbf{H}_a \cdot \mathbf{H}_b - \mathbf{E}_a \cdot \mathbf{E}_b \right) \left( \mathbf{H}_c \cdot \mathbf{H}_d - \mathbf{E}_c \cdot \mathbf{E}_d \right) \\\nonumber
    &+& \left[ \frac{\lambda_4}{m^4} f_{abc} f_{ade} + \frac{\lambda_6}{m^4} k_{bcde} \right] \left( \left( \mathbf{E}_b \cdot \mathbf{E}_d - \mathbf{H}_b \cdot \mathbf{H}_d \right) \left( \mathbf{E}_c \cdot \mathbf{E}_e - \mathbf{H}_c \cdot \mathbf{H}_e \right) + \left( \mathbf{E}_b \cdot \mathbf{E}_c - \mathbf{H}_b \cdot \mathbf{H}_c \right) \left( \mathbf{E}_d \cdot \mathbf{E}_e - \mathbf{H}_d \cdot \mathbf{H}_e \right) \right.\\\nonumber
    &+&\left. \left(\mathbf{E}_b \cdot \mathbf{H}_e\right) \left( \mathbf{E}_c \cdot \mathbf{H}_d + \mathbf{H}_c \cdot \mathbf{E}_d \right) + \left( \mathbf{H}_b \cdot \mathbf{E}_e \right) \left( \mathbf{H}_d \cdot \mathbf{E}_c \right) \right) 
    +  \frac{4\lambda_5}{m^4} f_{abc} f_{ade} \left( \mathbf{E}_b \cdot \mathbf{E}_d - \mathbf{H}_b \cdot \mathbf{H}_d \right) \left( \mathbf{E}_c \cdot \mathbf{E}_e - \mathbf{H}_c \cdot \mathbf{H}_e \right)  \\ \nonumber
    &+& \left[ \frac{\lambda_7}{m^4} f_{abc} + \frac{\lambda_8}{m^4} d_{abc} \right] \left( \left( \mathcal{D}_t \rho_a \right) \mathbf{E}_b \cdot \mathbf{E}_c - \left( \mathcal{D}_t \mathbf{j}_a \right) \cdot \mathbf{E}_b \times \mathbf{H}_c -
     \left( \mathbf{E}_b \cdot \boldsymbol{\mathcal{D}} \;
    \mathbf{j}_a   \cdot \right) \mathbf{E}_c  + \left( \boldsymbol{\mathcal{D}} \rho_a \right) \cdot \left( \mathbf{H}_b \times \mathbf{E}_c \right) \right. \\\nonumber
    &+& \left. \left( \boldsymbol{\mathcal{D}} \cdot \mathbf{j}_a \right) \mathbf{H}_b \cdot \mathbf{H}_c - \left( \mathbf{H}_c \boldsymbol{\mathcal{D}} j^i_a \right) H^i_b \right)
    + \frac{\lambda_9}{m^4}f_{abc} \left( 2 \mathbf{E}_a \cdot \mathbf{j}_b \rho_c + \mathbf{H}_a \cdot \left( \mathbf{j}_b \times \mathbf{j}_c \right) \right) 
    \\ 
    &+& \left. 
    \frac{\lambda_2}{m^2} \left( (\rho_a)^2 - \mathbf{j}_a \cdot \mathbf{j}_a \right)
    +      \frac{\lambda_{10}}{m^4} \left[
    ({\cal D}_t \rho^a)^2
    + (\boldsymbol{ {\cal D}}\cdot \mathbf{j}^a)^2 
    - (\boldsymbol{ {\cal D}}  \rho^a)^2
    -  ({\cal D}_t  \;\mathbf{j}^a)^2
    \right]  \right\}.
\end{eqnarray}

For the terms exclusively dependent on the c-electromagnetic fields a wide variety of couplings show up, most of them carrying the non-Abelian character either in the c-electric, c-magnetic sectors or mixed. 
The original results of this work, however, are present in the gluon couplings to quark currents.
The imaginary part ($\lambda_8$)
contains several quark current or quark density interactions with c-electric and/or c-magnetic fields.
Some of the terms present the structure of the (Abelian) Poynting vector that is present in the non-Abelian  one, as  shown below.
 Although these terms may be related to pair creation, 
they may 
be possibly  identified to 
emission or absorption of a gluon-wave 
or a
related effect that can include attenuation.
It is also interesting to notice that the imaginary part 
 ($\lambda_8$) does not disappear in the 
 pure c-magnetic sector. 
 To produce an entirely real effective action in the presence of matter fields, one needs further restrictions:
 \begin{eqnarray}
 \label{cond-im=0}
{\cal I}m (\Gamma_q) = 0  &\;\;\;
\to &
\;\;\;\;\;\;\;\;
d_{abc} \left\{  [ (\mathbf{H}_c \cdot \boldsymbol{ {\cal D}})\mathbf{j}^a ] \cdot 
 \mathbf{H}^b
    - (\boldsymbol{ {\cal D}} \cdot \mathbf{j}^a)
    \mathbf{H}^b \cdot \mathbf{H}^c \right\} 
    = 0,
    \;\;\;\;\;\;\;\;
    \mathbf{E}_a = 0,
    \nonumber 
    \\
{\cal I}m (\Gamma_q) = 0  &\;\;\;
\to &
\;\;\;\;\;\;\;\;
d_{abc} \left\{ \left( \mathcal{D}_t \rho_a \right) \mathbf{E}_b \cdot \mathbf{E}_c -   \left( \mathbf{E}_b \cdot \boldsymbol{\mathcal{D}} \;
    \mathbf{j}_a   \cdot \right) \mathbf{E}_c   \right\} =  0,
\;\;\;\;\;\;\;\;\;\;
\;\;\;\;
\mathbf{B}_a = 0,
\nonumber
\\
&& \mbox{Or mixed conditions for}
\;\;\;\;
\mathbf{E}_a , \mathbf{B}_a \neq 0.
\end{eqnarray}
In general 
time and spatial  covariant derivatives
cannot be reduced to simple derivatives due to contributions of   gluon field,
being consequently higher order in the chromo-electromagnetic field.
Except for these terms  in  the covariant derivatives, one 
may have real effective action for 
time -independent 
and homogeneous quark densities and currents.
With these conditions, in the vacuum or around
quark sources, the reality of the action with purely c-magnetic fields  (c-electric fields), 
is ensured at this order  in the expansion of the  quark determinant.
Besides that, there are  very different structures corresponding to quark-antiquark-gluon interactions 
in the c-electric, and c-magnetic sectors and for both.
These terms, except for 
$\lambda_2$ and $\lambda_9$, depend necessarily on time-dependent and/or spatial non-uniform quark densities or currents.
For uniform quark sources, two terms survive: $\lambda_2$ and 
non-Abelian  terms from $\lambda_9$.

\subsection{   Energy momentum tensor
and its Trace
}

The symmetrized energy-momentum tensor (EMT)  can be obtained in a gauge-invariant form, as proposed in \cite{Blaschke2016}. To do so, firstly we decompose the effective Lagrangian into:
\begin{align}
    \mathcal{L} = \mathcal{L}_I(G, J, \mathcal{D}J, G \mathcal{D}J) + \mathcal{L}_D(D\psi) + \mathcal{L}_0(\psi),
\end{align}
where  
Lagrangian terms were split into  
the QCD-EH interaction part ($\mathcal{L}_I$), a Dirac (kinetic) part ($\mathcal{L}_D$), and a pure fermion part ($\mathcal{L}_0$)  simply given by the 
self interacting term $J^2$ and 
the mass terms in equation \eqref{Seffresult}. 
By considering 
that only $\mathcal{L}_I$ depends on derivatives of $A_{\mu a}$,
the canonical EMT becomes :
\begin{align} 
    T_{\mbox{\tiny can}}^{\mu \nu} =& \frac{\partial \mathcal{L}_I}{\partial (\partial_\mu A_{\rho a})} \partial^{\nu} A_{\rho a}  + \partial^\nu \Bar{\psi} \frac{\partial \mathcal{L}_D}{\partial (\partial_\mu \Bar{\psi})} + \frac{\partial \mathcal{L}_D}{\partial (\partial_\mu {\psi})}\partial^\nu {\psi} 
     + \partial^\nu \Bar{\psi} \frac{\partial \mathcal{L}_I}{\partial (\partial_\mu \Bar{\psi})} + \frac{\partial \mathcal{L}_I}{\partial (\partial_\mu {\psi})}\partial^\nu {\psi} - g^{\mu \nu} (\mathcal{L}_I +\mathcal{L}_D + \mathcal{L}_0) .
     \label{EMT}
\end{align}
Since $\mathcal{L}_I$ depends on $\partial_\mu A_{\nu a}$ only trough $G_{\mu \nu a}$, we can do a variable change and use the definition $G_{\mu \rho a} = \partial_\mu A_{\rho a} - \partial_\rho A_{\mu a} + g f_{abc} A_{\mu b} A_{\rho c}$ in the first term of Eq. \eqref{EMT}:
\begin{align} 
    \frac{\partial \mathcal{L}_I}{\partial (\partial_\mu A_{\rho a})} \partial^{\nu} A_{\rho a} =& 2 \frac{\partial \mathcal{L}_I}{\partial G_{\mu \rho a}} G_{\mbox{ } \rho a}^{\nu}  + 2 \partial_\rho \left( \frac{\partial \mathcal{L}_I}{\partial G_{\mu \rho a}}  A^\nu_{ a} \right) +  \left( \partial_\rho  \frac{\partial \mathcal{L}_I}{\partial(\partial_\rho A_{\mu a})}  \right) A^\nu_{ a} - 2 gf_{abc} \frac{\partial \mathcal{L}_I}{\partial G_{\mu \rho a}} A^\nu_b A_{\rho c}.
\end{align}

The second term is of the form $-\partial_\rho \chi^{\mu \rho \nu}$, with $\chi^{\mu \rho \nu} = - \chi^{\rho \mu \nu}$. Since the EMT can be redefined by adding such a function without any loss \cite{Peskin2018}, we disregard this term from now on, and call the EMT simply by $T^{\mu \nu}$, since it is not the canonical EMT anymore. 
Using the tree-level Euler-Lagrange equations, without the EH terms
in the expression above, it yields:
\begin{align} 
    \frac{\partial \mathcal{L}_I}{\partial (\partial_\mu A_{\rho a})} \partial^{\nu} A_{\rho a} 
    %= & 2 \frac{\partial \mathcal{L}_I}
    %{\partial G_{\mu \rho a}} G_{\mbox{ } 
    %\rho a}^{\nu}  +  \left( \frac{\partial 
    %\mathcal{L}_I}{\partial A_{\mu a}} + 
    %\frac{\partial \mathcal{L}_D}{\partial 
    %A_{\mu a}}  \right) A^\nu_{ a} - 2 
    %gf_{abc} \frac{\partial \mathcal{L}_I}
    %{\partial G_{\mu \rho a}} A^\nu_b 
    %A_{\rho c} \\
    \label{dois termos zeram}
        = & 2 \frac{\partial \mathcal{L}_I}{\partial G_{\mu \rho a}} G_{\mbox{ } \rho a}^{\nu} + \frac{\partial \mathcal{L}_D}{\partial A_{\mu a}} A^\nu_{ a} - 2 gf_{abc} \frac{\partial \mathcal{L}_I}{\partial G_{\mu \rho a}} A^\nu_b A_{\rho c} + \frac{\partial \mathcal{L}_I}{\partial A_{\mu a}} A^\nu_a.
\end{align}

The last two terms of equation \eqref{dois termos zeram} must be treated carefully. The derivative with respect to $A_{\mu a}$ can come from two types of terms:
\begin{itemize}
    \item The ones that contain the field $G_{\mu \nu a}$;
    \item The ones that contain the covariant derivative $\mathcal{D}_\mu$.
\end{itemize}

With that in mind, we make the definition:
\begin{align}
    \frac{\partial}{ \partial A_{\mu a}} = \frac{\partial'}{\partial' A_{\mu a}} + \frac{\partial''}{\partial'' A_{\mu a}},
\end{align}
where the derivative with prime means deriving with respect to $A_{\mu a}$ from the $G_{\mu \nu a}$ field and the derivative with two primes is deriving with respect to $A_{\mu a}$ from the $\mathcal{D}_\mu$. With these, the last two terms in Eq. \eqref{dois termos zeram}:
\begin{align}\nonumber
    I^{\mu \nu} \equiv & - 2 gf_{abc} \frac{\partial \mathcal{L}_I}{\partial G_{\mu \rho a}} A^\nu_b A_{\rho c} + \frac{\partial \mathcal{L}_I}{\partial A_{\mu a}}A^\nu_a \\\nonumber
     =& - 2 gf_{abc} \frac{\partial \mathcal{L}_I}{\partial G_{\mu \rho a}} A^\nu_b A_{\rho c} + \frac{\partial' \mathcal{L}_I}{\partial' A_{\mu a}}A^\nu_a + \frac{\partial'' \mathcal{L}_I}{\partial'' A_{\mu a}}A^\nu_a \\\nonumber
    =& - 2 gf_{abc} \frac{\partial \mathcal{L}_I}{\partial G_{\mu \rho a}} A^\nu_b A_{\rho c} + \frac{\partial' \mathcal{L}_I}{\partial' G_{\alpha \beta b}} \frac{\partial' G_{\alpha \beta b}}{\partial' A_{\mu a}} A^\nu_a + \frac{\partial'' \mathcal{L}_I}{\partial'' A_{\mu a}}A^\nu_a\\\nonumber
    =& - 2 gf_{abc} \frac{\partial \mathcal{L}_I}{\partial G_{\mu \rho a}} A^\nu_b A_{\rho c} + \frac{\partial \mathcal{L}_I}{\partial G_{\alpha \beta b}} gf_{bcd} \left( \delta_{ca} g_{\alpha}^\mu A_{\beta d} + \delta_{da} g_{\beta}^{\mu} A_{\alpha c} \right) A^\nu_a + \frac{\partial'' \mathcal{L}_I}{\partial'' A_{\mu a}}A^\nu_a \\\nonumber
    =&- 2 gf_{abc} \frac{\partial \mathcal{L}_I}{\partial G_{\mu \rho a}} A^\nu_b A_{\rho c} + gf_{bad} \frac{\partial \mathcal{L}_I}{\partial G_{\mu \beta b}} A_{\beta d}A^\nu_a + gf_{bca} \frac{\partial \mathcal{L}_I}{\partial G_{\alpha \mu b}} A_{\alpha c}A^\nu_a + \frac{\partial'' \mathcal{L}_I}{\partial'' A_{\mu a}}A^\nu_a\\\nonumber
    =& - 2 gf_{abc} \frac{\partial \mathcal{L}_I}{\partial G_{\mu \rho a}} A^\nu_b A_{\rho c} + gf_{abc} \frac{\partial \mathcal{L}_I}{\partial G_{\mu \rho a}} A_{\rho c} A^\nu_b + gf_{acb} \frac{\partial \mathcal{L}_I}{\partial G_{\rho \mu a}} A_{\rho c} A^\nu_b + \frac{\partial'' \mathcal{L}_I}{\partial'' A_{\mu a}}A^\nu_a\\
    =& \frac{\partial'' \mathcal{L}_I}{\partial'' A_{\mu a}}A^\nu_a.
\end{align}

By writing  all the terms in eq. \eqref{EMT}:
\begin{align}\nonumber
    T^{\mu \nu} =& 2 \frac{\partial \mathcal{L}_I}{\partial G_{\mu \rho a}} G_{\mbox{ } \rho a}^{\nu}  + \frac{\partial \mathcal{L}_D}{\partial A_{\mu a}} A^\nu_{ a}  + \partial^\nu \Bar{\psi} \frac{\partial \mathcal{L}_D}{\partial (\partial_\mu \Bar{\psi})} + \frac{\partial \mathcal{L}_D}{\partial (\partial_\mu {\psi})}\partial^\nu {\psi} 
   +  \frac{\partial'' \mathcal{L}_I}{\partial'' A_{\mu a}}A^\nu_a + \partial^\nu \Bar{\psi} \frac{\partial \mathcal{L}_I}{\partial (\partial_\mu \Bar{\psi})} + \frac{\partial \mathcal{L}_I}{\partial (\partial_\mu {\psi})}\partial^\nu {\psi} 
    \\
    &
    - g^{\mu \nu} (\mathcal{L}_I +\mathcal{L}_D + \mathcal{L}_0).
\end{align}
By rearranging these terms, it follows:
\begin{align}\label{EMTdecomp1}
    T^{\mu \nu} &= T^{\mu \nu}_I + T^{\mu \nu}_D + T^{\mu \nu}_c - g^{\mu \nu} \mathcal{L}_0,
\end{align}
where:
\begin{align}
    T^{\mu \nu}_I =& 2 \frac{\partial \mathcal{L}_I}{\partial G_{\mu \rho a}} G_{\mbox{ } \rho a}^{\nu} - g^{\mu \nu} \mathcal{L}_I;\\
    T^{\mu \nu}_D =& \frac{\partial \mathcal{L}_D}{ \partial A_{\mu a}} A^\nu_a + \partial^\nu \Bar{\psi} \frac{\partial \mathcal{L}_D}{\partial (\partial_\mu \Bar{\psi})} + \frac{\partial \mathcal{L}_D}{\partial (\partial_\mu {\psi})}\partial^\nu {\psi} - g^{\mu \nu} \mathcal{L}_D;\\
    T^{\mu \nu}_c=& \frac{\partial'' \mathcal{L}_I}{ \partial'' A_{\mu a}} A^\nu_a + \partial^\nu \Bar{\psi} \frac{\partial \mathcal{L}_I}{\partial (\partial_\mu \Bar{\psi})} + \frac{\partial \mathcal{L}_I}{\partial (\partial_\mu {\psi})}\partial^\nu {\psi}.
\end{align}

By using Eq. \eqref{EMTdecomp1}, we calculate the complete symmetrized energy-momentum tensor  
and we obtain the following
equation:
\begin{align}
\nonumber
    T^{\mu \nu} =& - \frac{\lambda^2}{\lambda_R^2} G^{\mu \xi}_z G^\nu_{\mbox{ } \xi z} + 6\frac{\lambda_1}{m^2} f_{abc} G^{\nu \xi}_{a} G_{\xi \rho b} G^{\rho \mu}_{c} + 8\frac{\lambda_3}{m^4} k_{zbcd} G^{\mu \xi}_{b} G_{\alpha \beta c} G^{\alpha \beta}_{d} G^{\nu}_{\mbox{ } \xi z} \\\nonumber
    & + 4 \frac{\lambda_4}{m^4} f_{azc} f_{ade} \left( G^{\xi \zeta}_c G^{\mu \beta}_d + G^{\mu \beta}_c G^{\xi \zeta}_d \right) G_{\beta \zeta e} G^{\nu}_{\mbox{ } \xi z} + 8 \frac{\lambda_5}{m^4} f_{azc} f_{ade} G_{\alpha \beta c} G^{\mu \xi}_{d} G^{\alpha \beta}_{e} G^{\nu}_{\mbox{ } \xi z} \\\nonumber 
    & + 4 \frac{\lambda_6}{m^4} k_{zcde} \left( G^{\xi \zeta}_{c} G^{\mu \beta}_{d} + G^{\mu \beta}_c G^{\xi \zeta}_d \right) G_{\beta \zeta e} G^\nu_{\mbox{ } \xi z} + \frac{ \lambda_7 f_{azb} + \lambda_8 d_{azb}}{m^4} \left( \left\{ (\mathcal{D}^\mu J_\zeta)_a  - (\mathcal{D}_\zeta J^\mu)_a \right\}\right.\\\nonumber
    & \times G^{\xi \zeta}_c G^{\nu}_{\mbox{ } \xi b} - \left.\left\{ (\mathcal{D}^\xi J_\zeta)_a - (\mathcal{D}_\zeta J^\xi)_a \right\} G^{\mu \zeta}_c G^{\nu}_{\mbox{ } \xi b} + \frac{1}{2}(\mathcal{D}^\nu J^\beta)_a G^{\mu \alpha}_b G_{\alpha \beta c} + \frac{1}{2}(\mathcal{D}^\mu J^\beta)_a G^{\nu \alpha}_b G_{\alpha \beta c}  \right)  \\\nonumber
     & + \frac{i}{4} \left( \Bar{\psi} \gamma^\mu D^\nu \psi - (D^\nu \Bar{\psi}) \gamma^\mu \psi + (\mu \leftrightarrow \nu) \right) + 2 \frac{\lambda_9}{m^4} f_{zbc} J^\mu_b J^\xi_c G^{\nu}_{\mbox{ }\xi z} \\\nonumber
    &+ 2 \frac{\lambda_{10}}{m^4} (\mathcal{D}^\mu J^\beta)_a (\mathcal{D}^\nu J_\beta)_a - g^{\mu \nu} \left(-\frac{1}{4} \frac{g^2}{g_R^2} (G_a^{\mu_1 \nu_1})^2 + \Bar{\psi} \left( \slashed{D}-m \right) \psi + \frac{\lambda_1}{m^2} f_{abc} G_a^{\mu_1 \nu_1} G_{b \nu_1 \rho} G^{\rho}_{c \mbox{ } \mu}\right.
    \\
    \nonumber
    & + \frac{\lambda_2}{m^2} J^{\mu_1}_a J_{a \mu_1} + \frac{\lambda_3}{m^4} k_{abcd} G_{\mu_1 \nu_1 a} G^{\mu \nu}_b G_{\alpha \beta c} G^{\alpha \beta}_d + \frac{\lambda_4}{m^4} f_{abc} f_{ade} 
    G_{\mu_1 \alpha b} G^{\alpha \nu_1}_c G^{\mu_1 \beta}_{d} G_{\beta \nu_1 e} \\\nonumber
    & + \frac{\lambda_5}{m^4} f_{abc} f_{ade} G_{\mu_1 \nu_1 b} G_{\alpha \beta c} G^{\mu_1 \nu_1}_{d} G^{\alpha \beta}_e + \frac{\lambda_6}{m^4} k_{bcde} G_{\mu_1 \alpha b} G^{\alpha \nu_1}_c G^{\mu_1 \beta}_d G_{\beta \nu_1 e}  \\ \nonumber
    &  + \frac{\lambda_7}{m^4} f_{abc} (\mathcal{D}^{\mu_1} J_{\nu_1})_a G_{\mu_1 \alpha b} G^{\alpha \nu_1}_c + \frac{\lambda_8}{m^4} d_{abc} \left( \mathcal{D}^{\mu_1} J_{\nu_1} \right)_a G_{\mu_1 \alpha b} G^{\alpha \nu_1}_c \\\label{EMT}
    & \left. + \frac{\lambda_9}{m^4}f_{abc} G_{\mu_1 \nu_1 a} J^{\mu_1}_b J^{\nu_1}_c + \frac{\lambda_{10}}{m^4} \left(\mathcal{D}_\alpha J_{\mu_1} \right)_a \left(\mathcal{D}^\alpha J^{\mu_1}\right)_a \right),
\end{align}
where $(\mu \leftrightarrow \nu)$ means last expression, placed  before that symbol, with exchanged indices $\mu$ , $\nu$.

\subsubsection{ Trace of the EMT}
\label{traceanomaly}

The trace of the EMT
can then be written as:
\begin{align} 
\nonumber
    T^\mu_\mu =& 
    m \Bar{\psi} \psi + 2\frac{\lambda_1}{m^4}f_{abc} G^{\mu \xi}_{a} G_{\xi \rho b} G^{\rho \mu}_c - 4 \frac{\lambda_2}{m^2} (J^\mu_{a})^2 + 4 \frac{\lambda_3}{m^4}k_{abcd} G^{\mu \xi}_a G_{\mu \xi b} G^{\alpha \beta}_{c} G_{\alpha \beta d}  \\\nonumber
    & +4 \frac{\lambda_{4}}{m^4} f_{abc} f_{ade} G_{\mu \xi b} G^{\mu \beta}_c G^{\xi \zeta}_{d} G_{\beta \zeta e} +4 \frac{\lambda_5}{m^4}f_{abc}f_{ade} G_{\alpha \beta c} G^{\mu \xi}_{d} G^{\alpha \beta}_{e} G_{\mu \xi z} \\\nonumber
    &+ 4 \frac{\lambda_6}{m^4} k_{bcde} G_{\mu \xi b} G^{\xi \zeta}_c G^{\mu \beta}_{d} G_{\beta \zeta e} - \frac{\lambda_7 f_{azb} + \lambda_8 d_{azb}}{m^4}  (\mathcal{D}_{\zeta} J^\mu)_a G^{\xi \zeta}_b G_{\mu \xi z} \\
    \label{TRACEEMT}
    & - 2\frac{\lambda_9}{m^4} f_{abc} G_{\mu \nu a} J^{\mu}_b J^{\nu}_{c} - 3 \frac{\lambda_{10}}{m^4} ((\mathcal{D}_\mu J_\nu)_a)^2 .
\end{align}
 Scale invariance is  well known to be broken by fermion mass (the first term in the Eq. above) and  quantum gluon self interactions 
 \cite{Weinberg1995},
 all the other terms in this equation.
There are   terms from the 
quark-gluon interactions
$\lambda_7, \lambda_8$
and quark-quark interactions with  or without additional gluon insertions $\lambda_9, \lambda_{10}$ and $\lambda_2$.
The quark-gluon interaction terms contain both derivative dependent and independent  terms.
In the quark-gluon interaction sector, there are four terms ($\lambda_7, \lambda_8, \lambda_9$ and $\lambda_{10}$)
that correspond to higher order one or two gluon couplings to a quark current or quark self interactions.
There are two terms in the 
quark self interaction sector:
$\lambda_2$ and $\lambda_{10}$.
From the phenomenological point of view, these terms can be seen 
as   corrections to quark self interactions that are usually the basis of widely investigated models 
such as contact interaction models and  the Nambu-Jona-Lasinio model.
For instance, term $\lambda_{10}$ can be put in a more suitable form by means of an integration by parts
leading to $\mathcal{D}^2 J_\nu J^\nu \sim
m^2 J_\nu J^\nu+...$ for on shell quarks.
The interaction terms $\lambda_7, \lambda_8$ and $\lambda_9$ can contribute for quark self energy or quark or gluon self interactions
depending on particular gluon field configurations.
The  imaginary part of the effective action,
given by the term with $\lambda_8$,  
leads to  nontrivial  imaginary contributions
for quark-antiquark fields coupled to at least two gluons.
Overall, it  is seen  that these higher order interactions contribute   for the trace anomaly.

\subsection{
Energy and Poyinting vector in terms of Ec Bc}

From the EMT, we also extract
the energy and the Poyinting vector in terms of the chromo-electric and chromo-magnetic fields.
The energy 
can be written as:
\begin{align}\nonumber
    \epsilon \equiv T^{00} =& 
     - \Bar{\psi} \left( \slashed{D}-m \right) \psi
      + \frac{i}{2} \left( \Bar{\psi} \gamma^0 D_t \psi - (D_t \Bar{\psi}) \gamma^0 \psi \right)
     + \frac{\lambda^2}{2 \lambda_R^2} \left( \mathbf{H}_a \cdot \mathbf{H}_a + \mathbf{E}_a \cdot \mathbf{E}_a \right) + \frac{\lambda_1}{m^2} f_{abc} \left( 5 \mathbf{E}_a \times \mathbf{E}_b + \mathbf{H}_a \times \mathbf{H}_b  \right) \cdot \mathbf{H}_c 
     \\
     \nonumber
     & -4 \frac{\lambda_3}{m^4} k_{abcd} \left[ 2 (\mathbf{E}_a \cdot \mathbf{E}_b)(\mathbf{H}_c \cdot \mathbf{H}_d) - 3 (\mathbf{E}_a \cdot \mathbf{E}_b) (\mathbf{E}_c \cdot \mathbf{E}_d) + (\mathbf{H}_a \cdot \mathbf{H}_b) (\mathbf{H}_c \cdot \mathbf{H}_d) \right]  
    \\
    \nonumber
    & + \left( \frac{\lambda_4  f_{abc} f_{ade} 
    +  \lambda_6 k_{bcde} }{m^4}\right) \left[ 3 (\mathbf{E}_b \cdot \mathbf{E}_c) \left( \mathbf{E}_d \cdot \mathbf{E}_e - \mathbf{H}_d \cdot \mathbf{H}_e \right) +  (\mathbf{H}_b \cdot \mathbf{H}_c) \left( \mathbf{E}_d \cdot \mathbf{E}_e - \mathbf{H}_d \mathbf{H}_e \right) \right.\\\nonumber
    & + 3 (\mathbf{E}_b \cdot \mathbf{E}_d) \left(\mathbf{E}_c \cdot \mathbf{E}_e - \mathbf{H}_c \cdot \mathbf{H}_e \right)  + (\mathbf{H}_b \cdot \mathbf{H}_d) \left( \mathbf{E}_c \cdot \mathbf{E}_e - \mathbf{H}_c \cdot \mathbf{H}_e \right) + 3(\mathbf{E}_b \cdot \mathbf{H}_e) \left( \mathbf{E}_d \cdot \mathbf{H}_c + \mathbf{E}_c \cdot \mathbf{H}_d \right) \\\nonumber
    & \left.  - (\mathbf{H}_b \cdot \mathbf{E}_e) \left( \mathbf{E}_c \cdot \mathbf{H}_d + \mathbf{H}_c \cdot \mathbf{E}_d \right) \right] + 4\frac{\lambda_5}{m^4} f_{abc} f_{ade}\left[ 3 \mathbf{E}_d \cdot \mathbf{E}_b \left( \mathbf{E}_c \cdot \mathbf{E}_e - \mathbf{H}_c \cdot \mathbf{H}_e \right) + \mathbf{H}_b \cdot \mathbf{H}_d \left( \mathbf{E}_c \cdot \mathbf{E}_e - \mathbf{H}_c \cdot \mathbf{H}_e \right) \right] 
    \\
    \nonumber 
    & 
    + \frac{ \lambda_7 f_{azb} + \lambda_8 d_{azb}}{m^4}
    \left[ -2 \left(\mathbf{H}_c \times \mathcal{D}_t \mathbf{J}_a  \right) \cdot \mathbf{E}_b 
    + \left( \boldsymbol{\mathcal{D}} \cdot \mathbf{j}_a \right) \mathbf{H}_b \cdot \mathbf{H}_c - \left( \mathbf{H}_c \cdot \boldsymbol{\mathcal{D}} \mathbf{j}_a \right) \cdot \mathbf{H}_b  
    \right.\\\nonumber
    & \left.
    + \left( \mathbf{H}_c \times \boldsymbol{\mathcal{D}} \rho_a \right) \cdot \mathbf{E}_b - \left( \mathbf{H}_b \times \boldsymbol{\mathcal{D}} \rho_a \right) \cdot \mathbf{E}_c - 2 \left( \mathbf{E}_b \cdot \boldsymbol{\mathcal{D}} \mathbf{j}_a \right) \cdot \mathbf{E}_c + \left( \mathbf{E}_c \cdot \boldsymbol{\mathcal{D}} \mathbf{j}_a \right) \cdot \mathbf{E}_b + 2 \left( \mathcal{D}_t \rho_a \right) \mathbf{E}_b \cdot \mathbf{E}_c
    \right]
    \\
    \nonumber
    & 
    - \frac{\lambda_9}{m^4} 
    f_{abc} \left[ \mathbf{H}_a \cdot (\mathbf{j}_b \times \mathbf{j}_c)
    - \mathbf{E}_a \cdot (
    \rho_b \times \mathbf{j}_c) 
    + \mathbf{E}_a \cdot (
     \mathbf{j}_b \rho_c  ) 
    \right]
    \\\nonumber
    &+ \frac{\lambda_{10}}{m^4} \left[ \left( \mathcal{D}_t \rho_a  \right)^2 - \left( \mathcal{D}_t \mathbf{j}_a \right) \cdot \left( \mathcal{D}_t \mathbf{j}_a \right) + \left( \boldsymbol{\mathcal{D}} \rho_a \right) \cdot \left( \boldsymbol{\mathcal{D}} \rho_a \right) - \left( \boldsymbol{\mathcal{D}} j^j_a \right) \cdot \left( \boldsymbol{\mathcal{D}} j^j_a \right) \right] - \frac{\lambda_2}{m^2} \left( \rho_a^2 - \mathbf{j}_a \cdot \mathbf{j}_a \right) .
    \end{align}
 This equation contains static and time-dependent terms that survive or disappear in the limit of homogeneous quark sources and chromo-electromagnetic fields.
  This expression is intricate, with many possible contributions from different configurations.
  The imaginary part $\lambda_8$ 
 maybe set to zero for the following general conditions:
\begin{eqnarray}
0 &=& d_{azb} \left[ -2 \left(\mathbf{H}_c \times \mathcal{D}_t \mathbf{j}_a  \right) \cdot \mathbf{E}_b 
    + \left( \boldsymbol{\mathcal{D}} \cdot \mathbf{j}_a \right) \mathbf{H}_b \cdot \mathbf{H}_c - \left( \mathbf{H}_c \cdot \boldsymbol{\mathcal{D}} \mathbf{j}_a \right) \cdot \mathbf{H}_b  
    \right.\\\nonumber
    &+& \left.
    \left( \mathbf{H}_c \times \boldsymbol{\mathcal{D}} \rho_a \right) \cdot \mathbf{E}_b - \left( \mathbf{H}_b \times \boldsymbol{\mathcal{D}} \rho_a \right) \cdot \mathbf{E}_c - 2 \left( \mathbf{E}_b \cdot \boldsymbol{\mathcal{D}} \mathbf{j}_a \right) \cdot \mathbf{E}_c + \left( \mathbf{E}_c \cdot \boldsymbol{\mathcal{D}} \mathbf{j}_a \right) \cdot \mathbf{E}_b + 2 \left( \mathcal{D}_t \rho_a \right) \mathbf{E}_b \cdot \mathbf{E}_c
    \right].
\end{eqnarray}
Specific conditions for the real energy
in the 
c-magnetic and c-electric sectors 
are respectively given by:
\begin{eqnarray}
0 &=& d_{azb} \left[  \left( \boldsymbol{\mathcal{D}} \cdot \mathbf{j}_a \right) \mathbf{H}_b \cdot \mathbf{H}_c - \left( \mathbf{H}_c \cdot \boldsymbol{\mathcal{D}} \mathbf{j}_a \right) \cdot \mathbf{H}_b  
    \right], \;\;\;\;\;\;\;\;\;\;
    \mathbf{E}_b = 0,
    \nonumber
    \\
0 &=& d_{azb} \left[ - 2 \left( \mathbf{E}_b \cdot \boldsymbol{\mathcal{D}} \mathbf{j}_a \right) \cdot \mathbf{E}_c + \left( \mathbf{E}_c \cdot \boldsymbol{\mathcal{D}} \mathbf{j}_a \right) \cdot \mathbf{E}_b + 2 \left( \mathcal{D}_t \rho_a \right) \mathbf{E}_b \cdot \mathbf{E}_c
    \right], \;\;\;\;\;\; \mathbf{H}_b = 0.
\end{eqnarray}
Whereas the condition for the 
c-magnetic   sector 
(when $\mathbf{E}_b = 0$)
is the same as the condition
for real effective action
- as seen in Eq. \eqref{cond-im=0},
in the c-electric sector
($\mathbf{H}_b = 0.$) the 
conditions for these 
two quantities, to be real,  are different.

The Poynting vector, corresponding to
    the momentum,   
    can be written as:
\begin{align}    
\label{vector-Poyinting}
    \nonumber
    \mathbf{S} \equiv \sum_{i=1}^{3}T^{i0} \mathbf{e}_i =& \frac{\lambda^2}{\lambda_R^2} \mathbf{E}_a \times \mathbf{H}_a + 6\frac{\lambda_1}{m^2} f_{abc}\left( \mathbf{E}_a \cdot \mathbf{H}_c \right) \mathbf{H}_b + 16\frac{\lambda_3}{m^4} k_{abcd} \left( \mathbf{E}_a \times \mathbf{H}_b \right) \left( \mathbf{E}_c \cdot \mathbf{E}_d - \mathbf{H}_c \cdot \mathbf{H}_d \right)  \\\nonumber
    & + 4 \frac{\lambda_4}{m^4} f_{azc} f_{ade} \left( 2 \left(\left(\mathbf{E}_c \times \mathbf{E}_d \right) \cdot \mathbf{H}_e \right) \mathbf{E}_z + (\mathbf{H}_c \cdot \mathbf{H}_e - \mathbf{E}_c \cdot \mathbf{E}_e) \mathbf{E}_d \times \mathbf{H}_z - (\mathbf{H}_c \cdot \mathbf{E}_d + \mathbf{E}_c \cdot \mathbf{H}_d)  \mathbf{H}_e \times \mathbf{H}_z \right) \\\nonumber
    &+ 16 \frac{\lambda_5}{m^4} f_{azc} f_{ade} \mathbf{E}_z \times \mathbf{H}_d \left( \mathbf{E}_c \cdot \mathbf{E}_e - \mathbf{H}_c \cdot \mathbf{H}_e \right) \\\nonumber 
    & + 4 \frac{\lambda_6}{m^4} k_{zcde} \left( 2 \left[ \left( \mathbf{E}_c \times \mathbf{E}_d \right) \cdot \mathbf{H}_e \right] \mathbf{E}_z + (\mathbf{E}_d \times \mathbf{H}_z) \left( \mathbf{H}_c \cdot \mathbf{H}_e - \mathbf{E}_c \cdot \mathbf{E}_e \right) + (\mathbf{E}_c \times \mathbf{H}_z) \left( \mathbf{H}_d \cdot \mathbf{H}_e - \mathbf{E}_d \cdot \mathbf{E}_e \right) \right.\\\nonumber
    &\left. - \mathbf{H}_e \times \mathbf{H}_z \left( \mathbf{H}_c \cdot \mathbf{E}_d + \mathbf{E}_c \cdot \mathbf{H}_d \right) \right) \\\nonumber
    &+ \frac{ \lambda_7 f_{azb} + \lambda_8 d_{azb}}{m^4} \left( \left\{ (\mathbfcal{D} \rho)_a - (\mathcal{D}^0 \mathbf{j})_a \right\} \mathbf{E}_b \cdot \mathbf{E}_z + \left\{  (\mathbfcal{D} j^j)_a - (\mathcal{D}^j \mathbf{j})_a  \right\} \left( \mathbf{H}_b \times \mathbf{E}_z \right)^{j}  \right.\\\nonumber
    & \left.\left\{ \left((\mathbfcal{D} \rho)_a - (\mathcal{D}_t \mathbf{j})_a \right\} \cdot \mathbf{E}_z \right) \mathbf{E}_b + \left\{ (\mathcal{D}^j \mathbf{j})_a - (\mathbfcal{D} j^j)_a \right\} \times \mathbf{H}_b E^j_z \right)  \\\nonumber
     & + \frac{i}{4} \left( \Bar{\psi} \boldsymbol{\gamma} D_t \psi - (D_t \Bar{\psi}) \boldsymbol{\gamma} \psi + + \Bar{\psi} \gamma^0 \mathbf{D} \psi - (\mathbf{D} \Bar{\psi}) \gamma^0 \psi \right)\\\nonumber
    &+\frac{\lambda_7 f_{abc} + \lambda_8 d_{abc}}{2m^4}
    \left[ \left( (\mathcal{D}^t \mathbf{j})_a \cdot \mathbf{E}_c \right) \mathbf{E}_b + (\mathcal{D}^t \rho)_a \mathbf{E}_c \times \mathbf{H}_b - (\mathcal{D}^t \mathbf{j})_a \mathbf{H}_b \cdot \mathbf{H}_c + \left( (\mathcal{D}^t \mathbf{j})\cdot \mathbf{H}_b \right) \mathbf{H}_c \right.\\
    &\left. + (\mathbfcal{D} \rho)_a \mathbf{E}_b \cdot \mathbf{E}_c - (\mathbfcal{D} j^j)_a \left( \mathbf{E}_b \times \mathbf{H}_c \right)^j \right] \\\nonumber
    &
    - 2 \frac{\lambda_9}{m^4} f_{zbc} \mathbf{j}_b \left( \mathbf{j}_c \cdot \mathbf{E}_z \right) 
    + 2 \frac{\lambda_{10}}{m^4} \left( (\mathbfcal{D} \rho)_a (\mathcal{D}^t \rho)_a - (\mathbfcal{D} j^j)_a (\mathcal{D}^t j^j)_a \right).
\end{align}
All the structures analyzed above are present.
Gluon interactions carry
momenta in different ways in the
 non-Abelian generalization to the ones typical of the electromagnetic wave,
 as
seen for 
$\lambda_3, \lambda_4, \lambda_5,\lambda_6$.
Quark-gluon interactions also
may carry momentum for different configurations.
Among these terms, the imaginary contributions from $\lambda_8$ contain different structures  both  in the chromo-electric sector and  in the chromo-magnetic term with time-dependent quark sources.
Real Poynting vector is obtained, in general, with the following condition:
\begin{eqnarray}
 0 &= d_{abc}
    \left[ \left( (\mathcal{D}^t \mathbf{j})_a \cdot \mathbf{E}_c \right) \mathbf{E}_b + (\mathcal{D}^t \rho)_a \mathbf{E}_c \times \mathbf{H}_b - (\mathcal{D}^t \mathbf{j})_a \mathbf{H}_b \cdot \mathbf{H}_c + \left( (\mathcal{D}^t \mathbf{j})\cdot \mathbf{H}_b \right) \mathbf{H}_c \right.
    \nonumber
    \\
    &\left. + (\mathbfcal{D} \rho)_a \mathbf{E}_b \cdot \mathbf{E}_c - (\mathbfcal{D} j^j)_a \left( \mathbf{E}_b \times \mathbf{H}_c \right)^j \right].
\end{eqnarray}
Two terms are non-zero for c-electric and c-magnetic field mutually perpendicular, which can be associated 
with the Abelian Poyinting vector coupled to time-dependent quark density and spatial varying quark current.
These terms show radiation
absorption/emission may be switched on and off for suitable configurations.
However, other c-electric c-magnetic field configurations may  contribute to the complex Poynting vector and may correspond, for example,  to radiation attenuation.
 The specific conditions  for real vector Poynting in the 
 c-electric and c-magnetic sectors are the following:
 \begin{eqnarray}
 0 &=& d_{abc}
    \left[ \left( (\mathcal{D}^t \mathbf{j})_a \cdot \mathbf{E}_c \right) \mathbf{E}_b   + (\mathbfcal{D} \rho)_a \mathbf{E}_b \cdot \mathbf{E}_c  \right],
    \;\;\;\;\;\;\;\;\;\;\;
\mathbf{H}_c  = 0,    
\nonumber
\\
 0 &=& d_{abc}
    \left[   - (\mathcal{D}^t \mathbf{j})_a \mathbf{H}_b \cdot \mathbf{H}_c + \left( (\mathcal{D}^t \mathbf{j})\cdot \mathbf{H}_b \right) \mathbf{H}_c  \right],
       \;\;\;\;\;\;\;\;
\mathbf{E}_c  = 0.
 \end{eqnarray}
All these conditions, at this order in the 
c-electric and c-magnetic fields - i.e. by reducing the covariant derivative to simple derivatives - are related to 
time  and spatial (non)variation of the 
quark current and density. 
In addition to these dynamical quark terms,
there are higher-order non-homogeneous chromo-electric and chromo-magnetic fields.

\subsubsection{ Limit of constant fields}

To illustrate our results so far, we consider the background constant case:
\begin{align}\nonumber
    A^\mu_a =& \frac{1}{2} G^{\mu \nu} x_{\nu} n_a,
    \\
    \label{BackgroundConstant}
    j^\mu_a =& j^\mu n_a,
\end{align}
where $G^{\mu \nu}$ and $j^\mu$ are constants and $n_a$ is an unit vector in color space. 
By selecting the scalar sector of the quark current that may bring contribution for the vacuum,
a Fierz transformation can be performed and we pick up 
\begin{eqnarray}
j_{\mu a} j^\mu_a \sim  \frac{g^2}{2} \left( \Bar{\psi} \psi \right)^2 
\sim \frac{g^2 }{2} \langle  \Bar{\psi} \psi  \rangle^2,
\end{eqnarray}
where  
scalar quark current has been considered to 
generate the scalar quark-antiquark condensate 
 In this limit, the energy density and Poynting vector will be given by:
\begin{align} 
\label{Energy}
\nonumber
    \epsilon \left( \mathbf{H}, \mathbf{E}, \Bar{\psi}, \psi \right) =& + \frac{\lambda^2}{2 \lambda_R^2} \left( \mathbf{H} \cdot \mathbf{H} + \mathbf{E} \cdot \mathbf{E} \right) - 2 \frac{\lambda_3}{m^4} \left[ 2 (\mathbf{E} \cdot \mathbf{E})(\mathbf{H} \cdot \mathbf{H} - \mathbf{E} \cdot \mathbf{E}) - (\mathbf{E} \cdot \mathbf{E})^2 + (\mathbf{H} \cdot \mathbf{H})^2 \right]
    \\
    \nonumber
    & +  \frac{ \lambda_6 }{2m^4} \left[ 6 (\mathbf{E} \cdot \mathbf{E}) \left( \mathbf{E} \cdot \mathbf{E} - \mathbf{H} \cdot \mathbf{H} \right) +  2(\mathbf{H} \cdot \mathbf{H}) \left( \mathbf{E} \cdot \mathbf{E} - \mathbf{H} \cdot \mathbf{H} \right) + 4(\mathbf{E} \cdot \mathbf{H})^2 \right] 
    \\
    & + m \Bar{\psi} \psi - \frac{\lambda_2 g^2}{2m^2} \left( \Bar{\psi} \psi \right)^2 ,
    \\
    \mathbf{S} \left( \mathbf{H}, \mathbf{E}\right) =& \frac{\lambda^2}{\lambda_R^2} \mathbf{E} \times \mathbf{H}  + 4\frac{\left(2\lambda_3 +\lambda_6\right)}{m^4} \left( \mathbf{E} \times \mathbf{H} \right) \left( \mathbf{E} \cdot \mathbf{E} - \mathbf{H} \cdot \mathbf{H} \right).
\end{align}

\begin{figure}[H]
\centering
\begin{minipage}{.41\textwidth}
  \centering \includegraphics[width=.96\linewidth]{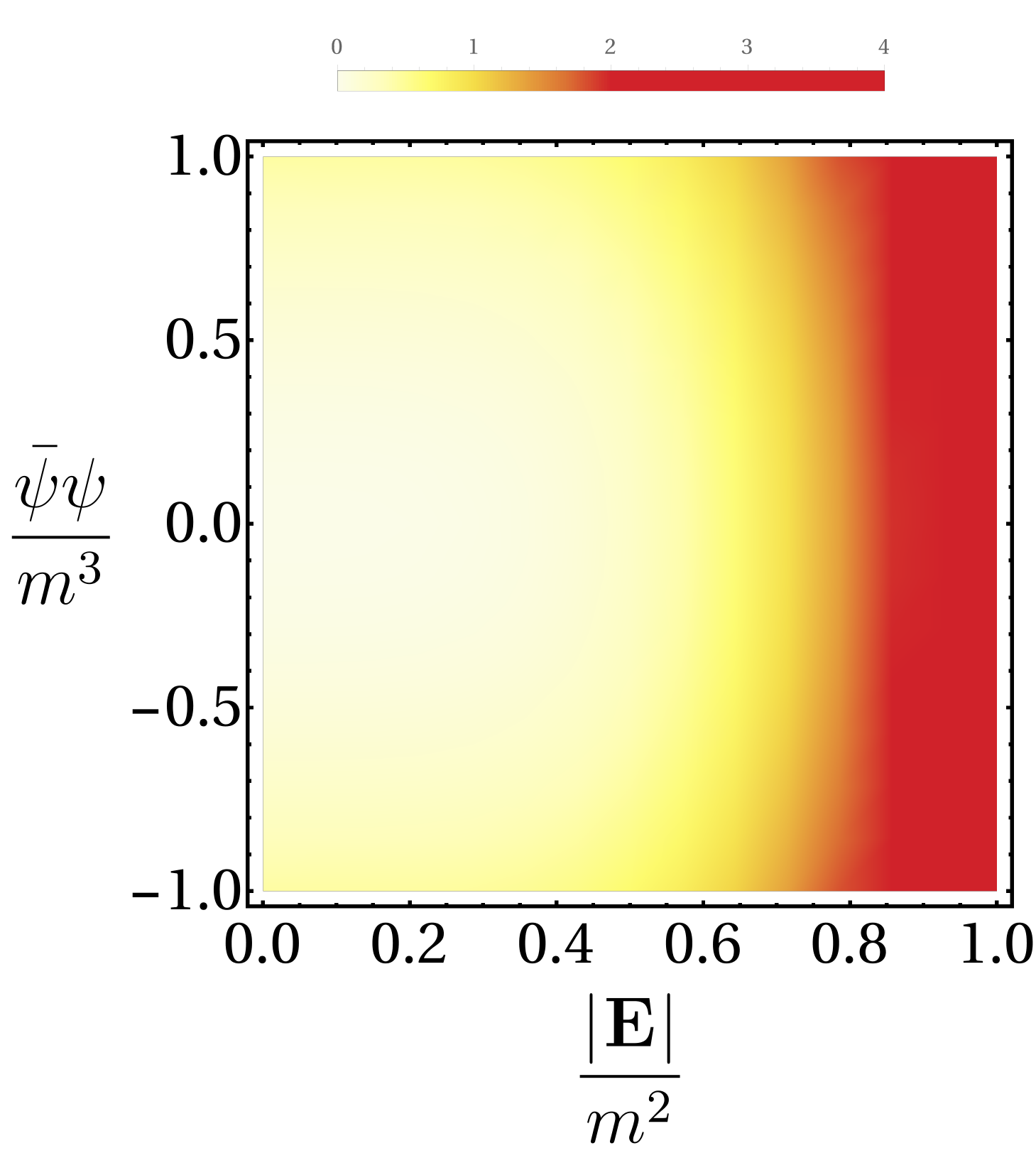}
  \caption{Heat-map of $2^7 3^2 \pi^2 \Delta \epsilon (\mathbf{H} = 0, \mathbf{E}, \Bar{\psi},\psi)$ as a function of the absolute value of the chromo-electric field, at energy scale 
$\Lambda = m$.  }
  \label{EnergyE}
\end{minipage}%
\hspace{1cm}
\begin{minipage}{.41\textwidth}
  \centering \includegraphics[width=.96\linewidth]{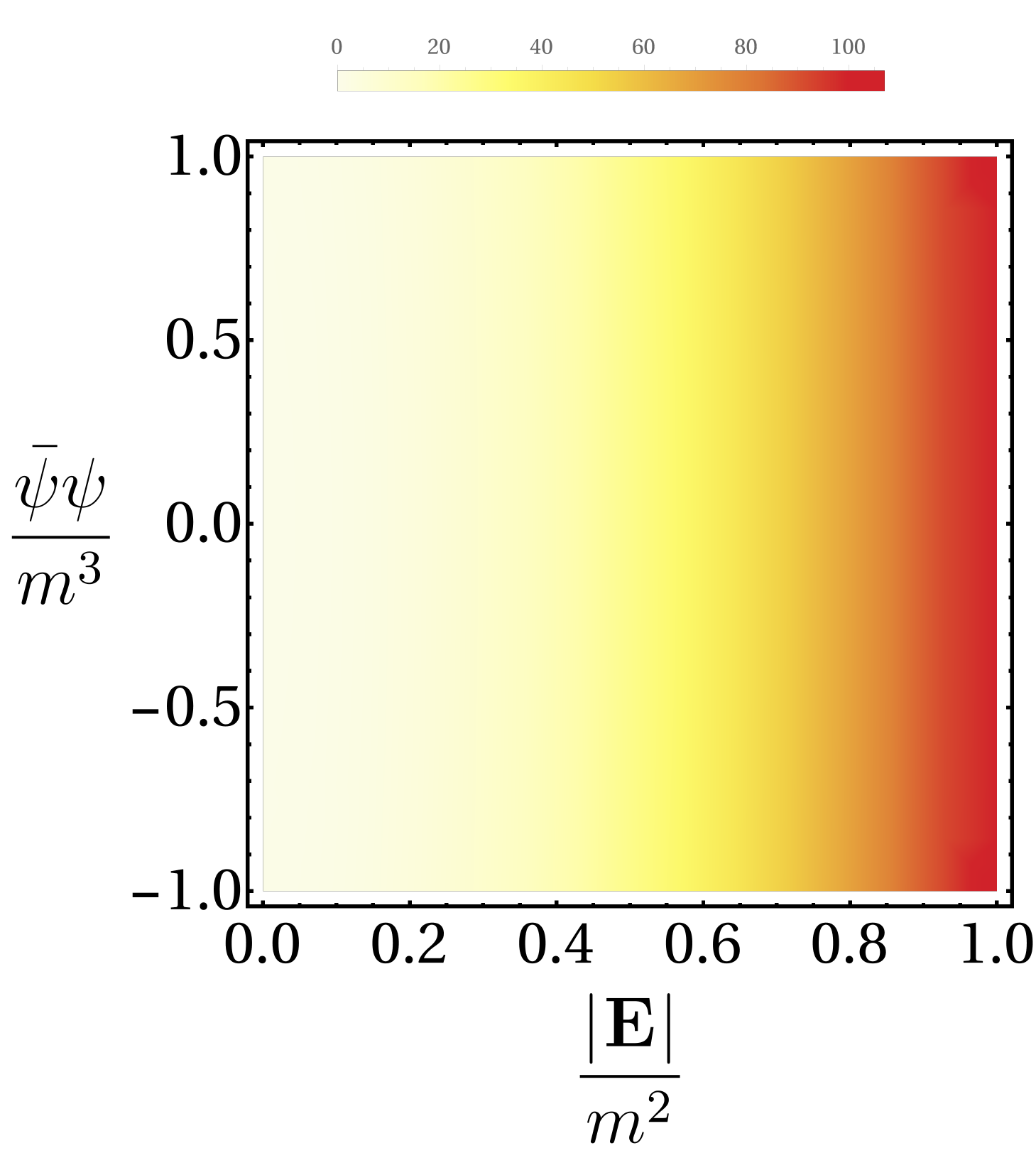}
  \caption{Heat-map of $2^7 3^2 \pi^2 \Delta \epsilon (\mathbf{H} = 0, \mathbf{E}, \Bar{\psi},\psi)$ as a function of the absolute value of the chromo-electric field, at energy scale $\Lambda = 3 m$.}
  \label{EnergyE3m}
\end{minipage}
\end{figure}

Note that, in this limit of 
constant fields, the Poynting vector does not depend on the matter fields. 
This can be expected, since only inhomogeneous quark currents should  carry momenta. 
Note that, by subtracting the correction to the classical 
energy density
($\epsilon_0 = \frac{1}{2} \left( \mathbf{H}^2 + \mathbf{E}^2 \right) + m\bar{\psi} \psi$)
we have the contribution exclusively due to the fermionic loop is then:
\begin{align} \label{DeltaEnergy}
\nonumber
    \Delta \epsilon (\mathbf{H}, \mathbf{E}, \Bar{\psi},\psi) 
    \equiv \epsilon - \epsilon_0 =& \frac{2 N_F g^2}{48 \pi^2} \ln{\left( \frac{\Lambda^2}{m^2} \right)} \left( \mathbf{H} \cdot \mathbf{H} + \mathbf{E} \cdot \mathbf{E} \right) - 2 \frac{\lambda_3}{m^4} \left[ 2 (\mathbf{E} \cdot \mathbf{E})(\mathbf{H} \cdot \mathbf{H} - \mathbf{E} \cdot \mathbf{E}) - (\mathbf{E} \cdot \mathbf{E})^2 + (\mathbf{H} \cdot \mathbf{H})^2 \right] 
\\\nonumber
    & +  \frac{ \lambda_6 }{2m^4} \left[ 6 (\mathbf{E} \cdot \mathbf{E}) \left( \mathbf{E} \cdot \mathbf{E} - \mathbf{H} \cdot \mathbf{H} \right) +  2(\mathbf{H} \cdot \mathbf{H}) \left( \mathbf{E} \cdot \mathbf{E} - \mathbf{H} \cdot \mathbf{H} \right) + 4(\mathbf{E} \cdot \mathbf{H})^2 \right] \\
    & - \frac{\lambda_2 g^2}{2m^2} \left( \Bar{\psi} \psi \right)^2.
\end{align} 
The profile of this correction to the energy density is given in Fig. \ref{EnergyE} for  $\Lambda = m$, that might 
be expected to be somewhat the situation 
 for  heavy quark masses,
 and in Fig. \ref{EnergyE3m} 
for  $\Lambda = 3 m$, that might be rather suitable
 lighter quarks 
  discussed around Eq. \eqref{convergence2}.
These heat-map figures are 
plotted as a functions of two quantities:
 scalar quark density (possibly associated to 
a scalar quark chiral condensate) 
$\bar{\psi} \psi$, and 
the c-electric field ($\mathbf{E}$) 
in the case of $\mathbf{H} = 0$.
Note that these contributions are always positive
possibly leading to a less stable system
and, most probably,
to more unstable vacuum.

\begin{figure}[H]
\centering
\begin{minipage}{.41\textwidth}
  \centering
  \includegraphics[width=.96\linewidth]{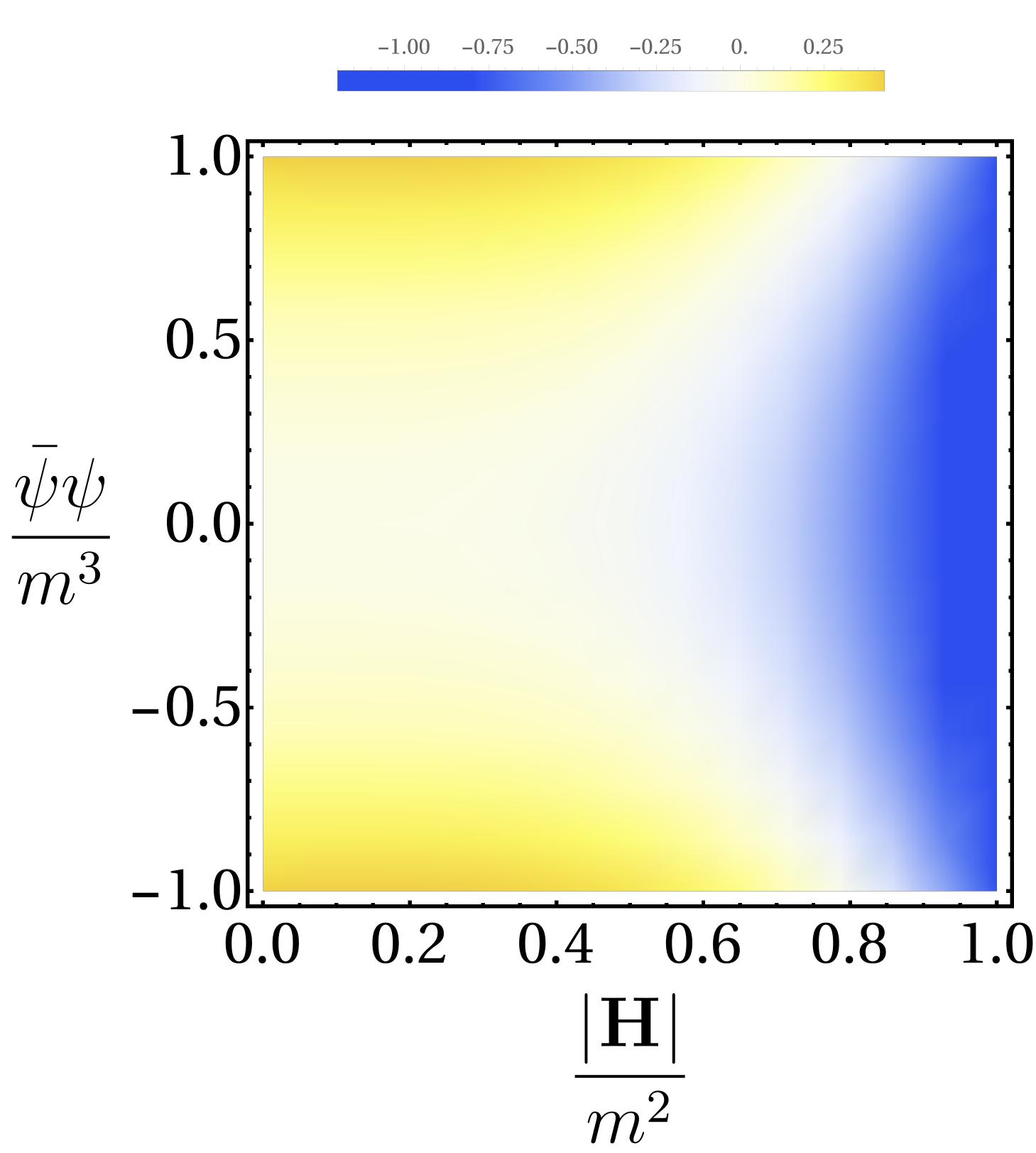}
  \caption{Heat-map of $2^7 3^2 \pi^2 \Delta \epsilon (\mathbf{H}, \mathbf{E}=0, \Bar{\psi},\psi)$ as a function of the absolute value of the chromo-magnetic field, at energy scale $\Lambda = m$.
}  \label{EnergyH}
\end{minipage}
\hspace{1cm}
\begin{minipage}{.41\textwidth}
  \centering
  \includegraphics[width=.96\linewidth]{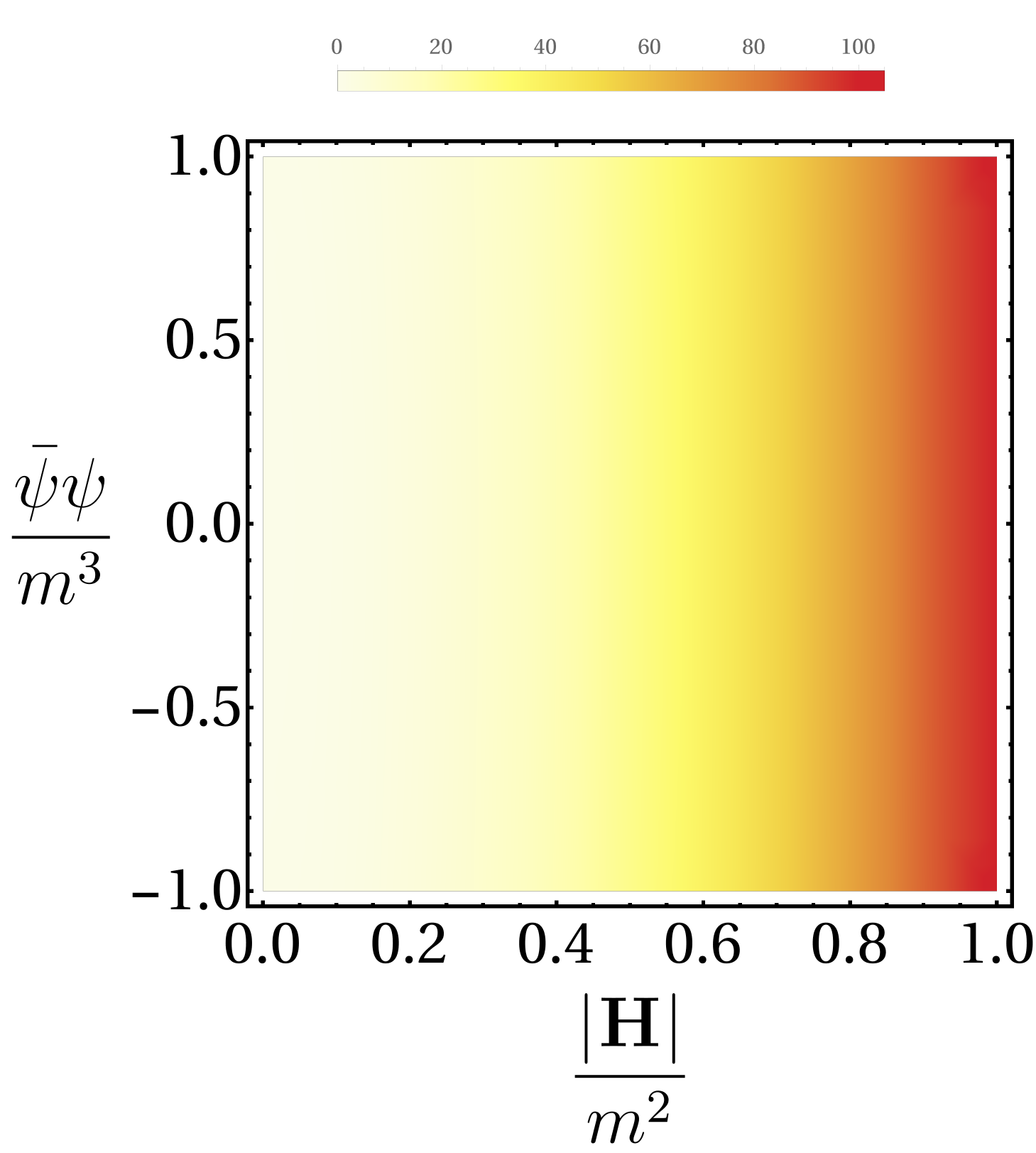}
  \caption{Heat-map of $2^7 3^2 \pi^2 \Delta \epsilon (\mathbf{H}, \mathbf{E}=0, \Bar{\psi},\psi)$ as a function of the absolute value of the chromo-magnetic field, at energy scale $\Lambda = 3m$.}
  \label{EnergyH3m}
\end{minipage}
\end{figure}

In Figs. \eqref{EnergyH} and \eqref{EnergyH3m}
the  correction to the energy density 
is plotted as function of the  same
scalar quark density, 
$\bar{\psi} \psi$, and 
the c-magnetic  field ($\mathbf{H}$) 
in the case of $\mathbf{E} = 0$
respectively for 
$\Lambda = m$ and $\Lambda = 3m$.
As discussed above, the first case can be expected to somehow
correspond to heavy quark contributions 
whereas the second figure to lighter quark contributions.
Although the increase of  
$\mathbf{H}$ can provide negative contributions for the 
energy density in the case of small cutoff (heavy quark 
picture) and then to more stable 
vacuum, it can lead to positive contributions
in the case of larger cutoff (light quark picture).

Note that Figs. \eqref{EnergyE3m} and \eqref{EnergyH3m}
look like each other and this is due to 
the dominance of the first terms in Eq. \eqref{DeltaEnergy}
for the large cutoff regime.
This dependence on the cutoff should manifest
in the renormalized energy density of Eq. \eqref{Energy}  by means of the 
particular renormalization scale.
Although the variation of the c-electric field leads, in case 
of small cutoff (or larger masses),  to larger variations of the energy density
than the variation of c-magnetic field, the dependence on the c-magnetic field can lead to a negative contribution to the energy density
for values of the quark scalar density close to
the  chiral quark-antiquark
scalar condensate \cite{PDG}. 
This might contribute to make the vacuum energy still more negative.
Several works indicate that the Yang-Mills vacuum is rather chromo-magnetic and the c-electric field destabilizes it \cite{Savvidy1977,Savvidy2020,Bali2014}.
The present estimation of contributions from the scalar quark antiquark condensate, in the case of somewhat heavy quarks
and cutoff $\lambda \sim m$,
 go along this picture.

\subsection{ Chromo-magnetic permeability and chromo-electric permittivity  }

We can also define the chromo-$\mathbf{B}$ and chromo-$\mathbf{D}$ fields,
and corresponding chromo-permeability and chromo-permittivity, from the effective action as \cite{Savvidy2020}:
\begin{align}
\label{B-H}
    \mathbf{B}_a =& -\eval{\frac{\partial \mathcal{L}}{\partial \mathbf{H}_a}}_{\mathbf{E}_a = 0},
    \\
    \label{D-E}
    \mathbf{D}_a =& + \eval{\frac{\partial \mathcal{L}}{\partial \mathbf{E}_a}}_{\mathbf{H}_a  = 0}.
\end{align}

In the case of vanishing sources, we can relate each of these fields with $\mu \mathbf{H}_a$ and $\epsilon  \mathbf{E}_a$ respectively, defining, therefore, corrections to the  chromo-permeability  and to 
the chromo-permittivity 
 due to the quark determinant. 
In the presence of sources, however, there will still appear reminiscent $\mathbf{B}_a$ and $\mathbf{D}_a$ fields even in the absence of  $\mathbf{H}_a$ and $\mathbf{E}_a$ fields due to the non-Abelian term $\lambda_9$.
Besides that, quark current and density may contribute to non-homogeneous and time-dependent configurations.
The resulting expressions for 
\eqref{B-H} and \eqref{D-E} are the following ones:
\begin{align}\nonumber
    \mathbf{B}_a =& \left( 1 + \frac{2 N_F g^2}{48 \pi^2} \ln{\frac{\Lambda^2}{m^2}} \right) \mathbf{H}_a + \frac{3\lambda_1}{m^2} f_{abc} \mathbf{H}_b \times \mathbf{H}_c - 16\frac{\lambda_3}{m^4} k_{abcd} \mathbf{H}_b \left( \mathbf{H}_c \cdot \mathbf{H}_d \right)\\\nonumber
    & - \frac{4\left( \lambda_4 + 4\lambda_5 \right)}{m^4} f_{abc} f_{dec} \mathbf{H}_d \left( \mathbf{H}_b \cdot \mathbf{H}_e \right) - \frac{4\lambda_6 }{m^4} k_{abcd} \left( \mathbf{H}_c \left( \mathbf{H}_b \cdot \mathbf{H}_d \right) + \mathbf{H}_b \left( \mathbf{H}_c \cdot \mathbf{H}_d \right) \right)\\\nonumber
    & + \frac{\lambda_7}{m^4} f_{abc} \left( \left( \boldsymbol{\mathcal{D}} j^i_b \right) H^i_c + \left(\mathbf{H}_c \cdot \boldsymbol{\mathcal{D}} \right) \mathbf{j}_b \right) + \frac{\lambda_8}{m^4} d_{abc} \left( \left( \boldsymbol{\mathcal{D}} j^i_b \right) H^i_c + \left(\mathbf{H}_c \cdot \boldsymbol{\mathcal{D}} \right) \mathbf{j}_b - 2 \left( \boldsymbol{\mathcal{D}} \cdot \mathbf{j}_b \right) \mathbf{H}_c \right)
    \\\label{B-H-Result}
    &- \frac{\lambda_9}{m^4} f_{abc} \mathbf{j}_b \times \mathbf{j}_c,
    \\
    \nonumber
    \mathbf{D}_a =& \left( 1 + \frac{2 N_F g^2}{48 \pi^2} \ln{\frac{\Lambda^2}{m^2}} \right) \mathbf{E}_a + 16\frac{\lambda_3 }{m^4} k_{abcd} \mathbf{E}_b \left( \mathbf{E}_c \cdot \mathbf{E}_d \right) + \frac{4\left( \lambda_4 + 4\lambda_5 \right)}{m^4}f_{abc}f_{dec}\left(\mathbf{E}_b \cdot \mathbf{E}_e\right) \mathbf{E}_d\\\nonumber
    & + \frac{4 \lambda_6}{m^4} k_{abcd} \left( \mathbf{E}_c \left( \mathbf{E}_b \cdot \mathbf{E}_d \right) + \mathbf{E}_b \left( \mathbf{E}_c \cdot \mathbf{E}_d \right) \right) + \frac{\lambda_7}{m^4} f_{abc} \left( \left( \boldsymbol{\mathcal{D}} j^l_b \right) E^l_c + \left( \mathbf{E}_b \cdot \boldsymbol{\mathcal{D}} \right) \mathbf{j}_c \right)\\\label{D-E-Result}
    & + \frac{\lambda_8}{m^4} d_{abc} \left( 2\left( \mathcal{D}_t \rho_b \right) \mathbf{E}_c - \left( \boldsymbol{\mathcal{D}}j^i_b \right) E^i_c + \left( \mathbf{E}_b \cdot \boldsymbol{\mathcal{D}} \right) \mathbf{j}_c \right)- \frac{2\lambda_9}{m^4} f_{abc} \mathbf{j}_b \rho_c.
\end{align}
These expressions have a strong non-linear character with 
anisotropic terms even in the absence of matter.
Some  anisotropic terms survive in
the Abelian limit in the absence and the presence of matter.

Conditions for zero imaginary parts are, respectively, given by:
\begin{eqnarray}
 \Im  (\mathbf{B}_a) = 0 \;\;\;\;\;
d_{abc} 
\left[ \left( \boldsymbol{\mathcal{D}} j^i_b \right) H^i_c + \left(\mathbf{H}_c \cdot \boldsymbol{\mathcal{D}} \right) \mathbf{j}_b - 2 \left( \boldsymbol{\mathcal{D}} \cdot \mathbf{j}_b \right) \mathbf{H}_c  
\right] &=& 0,
\nonumber
\\
 \Im  (\mathbf{D}_a) = 0 \;\;\;\;\;\;\;
d_{abc} \left[ 2\left( \mathcal{D}_t \rho_b \right) \mathbf{E}_c - \left( \boldsymbol{\mathcal{D}}j^i_b \right) E^i_c - \left( \mathbf{E}_b \cdot \boldsymbol{\mathcal{D}} \right) \mathbf{j}_c 
\right] &=& 0.
\end{eqnarray}
By replacing the covariant derivatives by simple derivatives, in general,
real observables can be obtained within the limit of time-independent charge density and spatially homogeneous currents. 
  Imaginary terms for 
the chromo-permittivity  (chromo-permeability)
are due to external chromo-magnetic (chromo-electric) fields and quark currents or time dependent density.
However, 
the symmetric structure constant, $d_{ijk}$, leads to particular combinations
 of the
chromoelectromagnetic fields and quark currents.

\subsubsection{ Limit of constant fields}

In the case of background constant fields, Eq. \eqref{BackgroundConstant}, we can write the fermionic loop correction to the $\mathbf{D}$ and $\mathbf{B}$ fields in Eq.'s \eqref{B-H-Result} and \eqref{D-E-Result} as:
\begin{align}\nonumber
    \boldsymbol{\Delta}_1 \equiv& \mathbf{B} - \mathbf{H} = \frac{2 N_F g^2}{48 \pi^2} \ln{\frac{\Lambda^2}{m^2}} \mathbf{H} - 8\frac{\lambda_3}{m^4} \mathbf{H} \left( \mathbf{H} \cdot \mathbf{H} \right) - 4\frac{\lambda_6 }{m^4} \mathbf{H} \left( \mathbf{H} \cdot \mathbf{H} \right),
    \\
    \nonumber
    \boldsymbol{\Delta}_2 \equiv& \mathbf{D} - \mathbf{E} = \frac{2 N_F g^2}{48 \pi^2} \ln{\frac{\Lambda^2}{m^2}} \mathbf{E} + 8\frac{\lambda_3 }{m^4} \mathbf{E} \left( \mathbf{E} \cdot \mathbf{E} \right) + 4\frac{ \lambda_6}{m^4}  \mathbf{E} \left( \mathbf{E} \cdot \mathbf{E} \right).
\end{align}

These functions are related to the chromo-electric and chromo-magnetic susceptibilities ($\chi_E$ and $\chi_M$ respectively) as $\boldsymbol{\Delta}_1 = \chi_E\mathbf{E}$ and $\boldsymbol{\Delta}_2 = \chi_M \mathbf{H}$ respectively.

\begin{figure}[H]
\centering
\begin{minipage}{.41\textwidth}
  \centering
  \includegraphics[width=.96\linewidth]{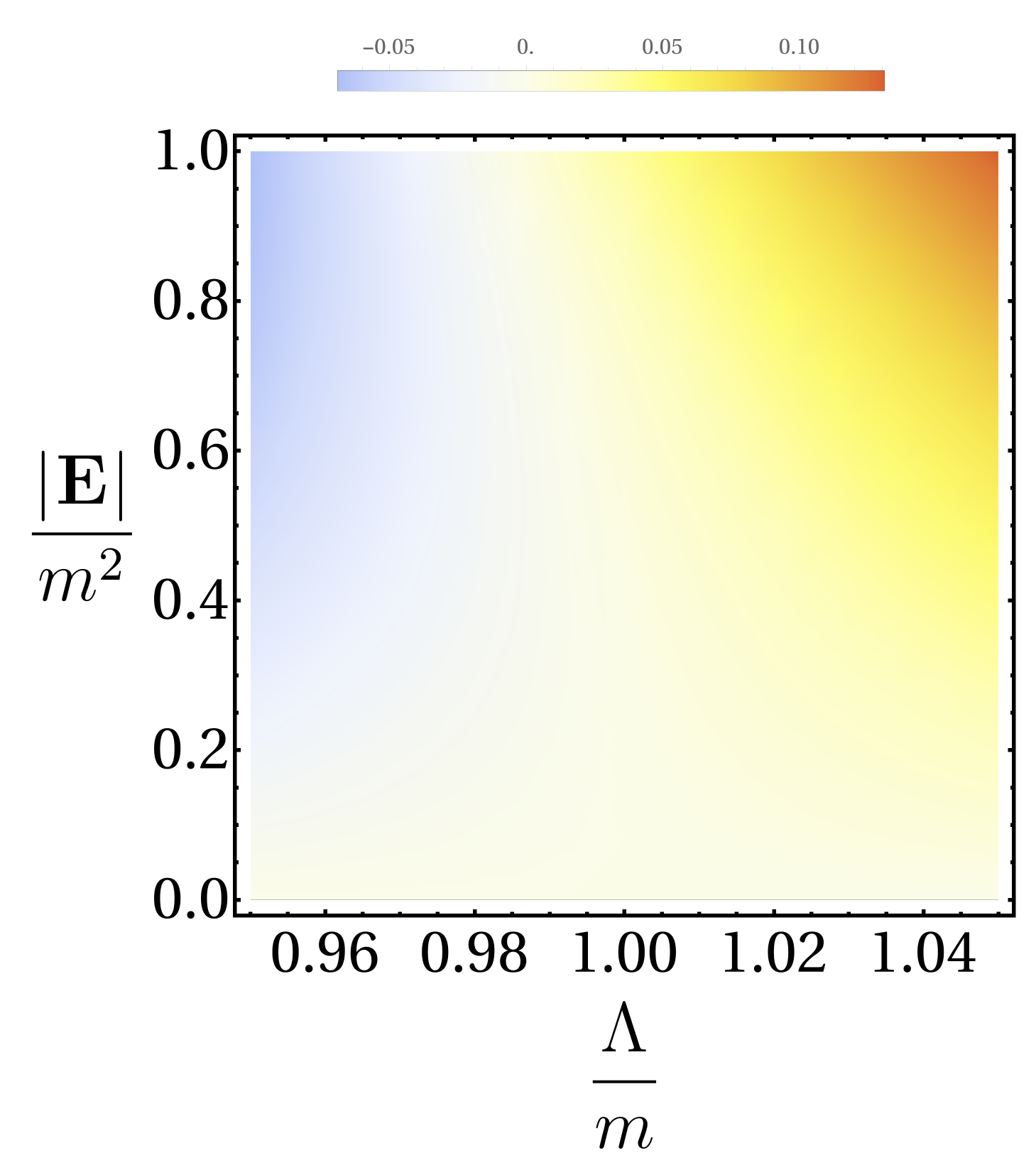}
  \caption{Heat-map of $2^3 3 \pi^2 \abs{\boldsymbol{\Delta}_2}$ as a function of the absolute value of the chromo-electric field and energy scale $\Lambda$ for the case of larger mass (smaller cutoff).}
  \label{DEFig}
\end{minipage}%
\hspace{1cm}
\centering
\begin{minipage}{.41\textwidth}
  \centering
  \includegraphics[width=.96\linewidth]{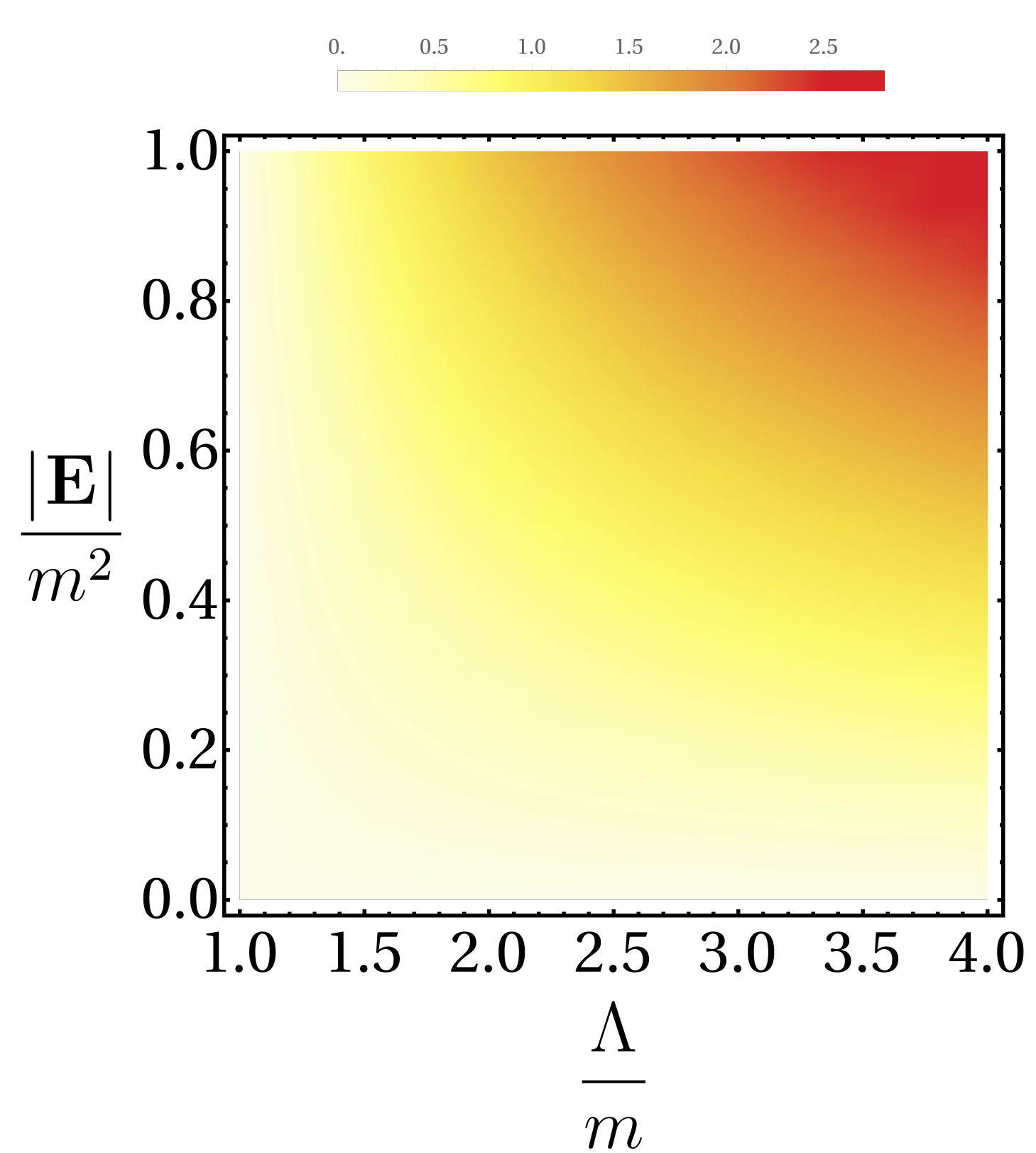}
  \caption{Heat-map of $2^3 3 \pi^2 \abs{\boldsymbol{\Delta}_2}$ as a function of the absolute value of the chromo-electric field and energy scale $\Lambda$.}
  \label{DEFig3m}
\end{minipage}%
\end{figure}

 In Figs. \ref{DEFig} and \ref{DEFig3m}
 we show the behavior of the
$|\Delta_2|$ (with $\mathbf{H}=0$)
 respectively
for a small range of the cutoff 
around $\Lambda \sim m$, as one might expect 
in the context of   heavy quarks,
and  for larger values of the cutoff $\Lambda >> m$,
that may be rather applied to light quark context.
The case of "small" cutoff presents a stronger variation
of the chromo-electric permittivity leading 
to possible negative values, although the variation
is quite small.
For larger values of the cutoff, these contributions 
are always positive.

\begin{figure}[H]
\begin{minipage}{.4\textwidth}
  \centering
  \includegraphics[width=.96\linewidth]{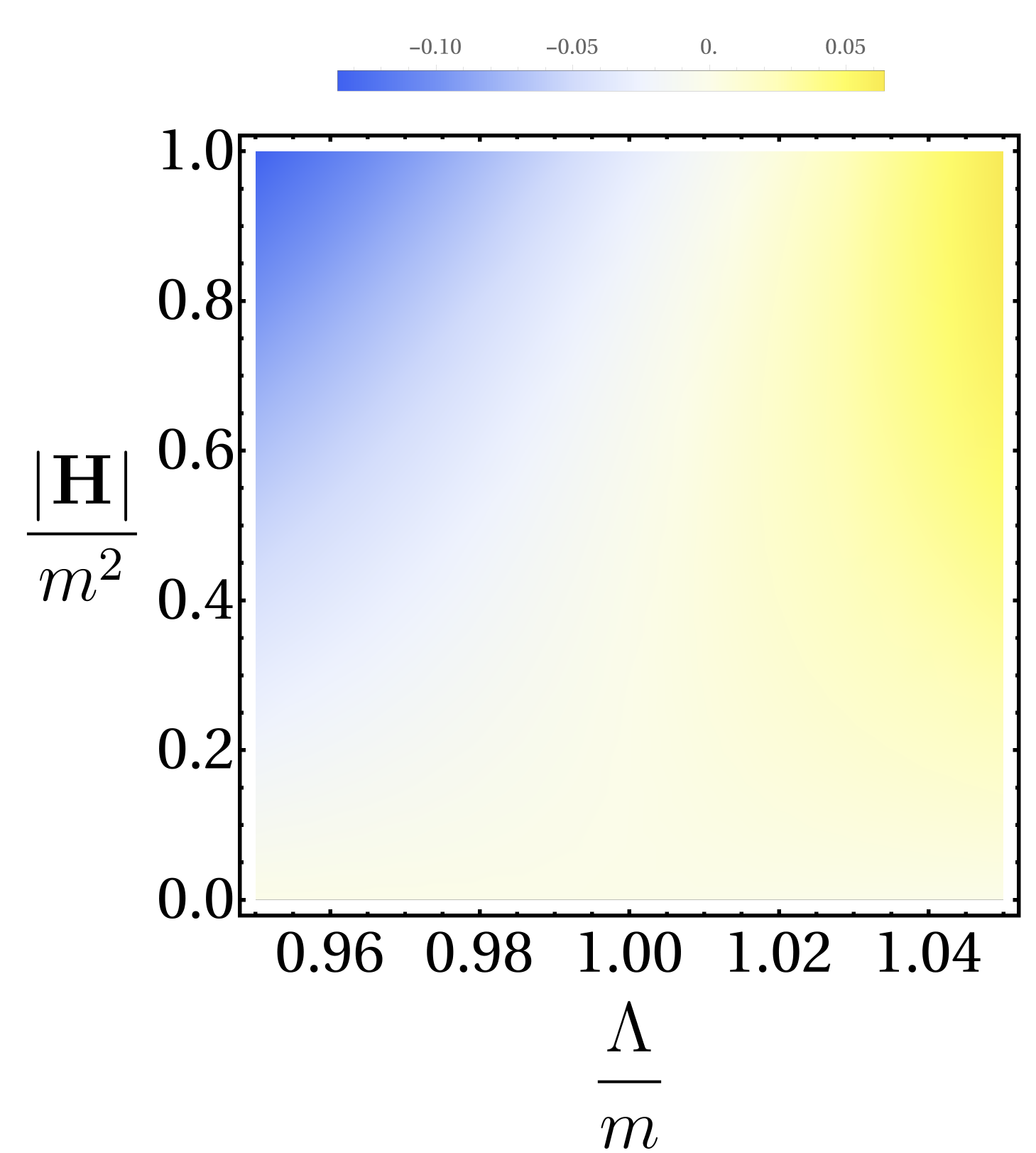}
  \caption{Heat-map of $2^3 3 \pi^2 \abs{\boldsymbol{\Delta}_1}$ as a function of the absolute value of the chromo-magnetic field and energy scale $\Lambda$.}
  \label{BHFig}
\end{minipage}
\hspace{1cm}
\begin{minipage}{.4\textwidth}
  \centering
  \includegraphics[width=.96\linewidth]{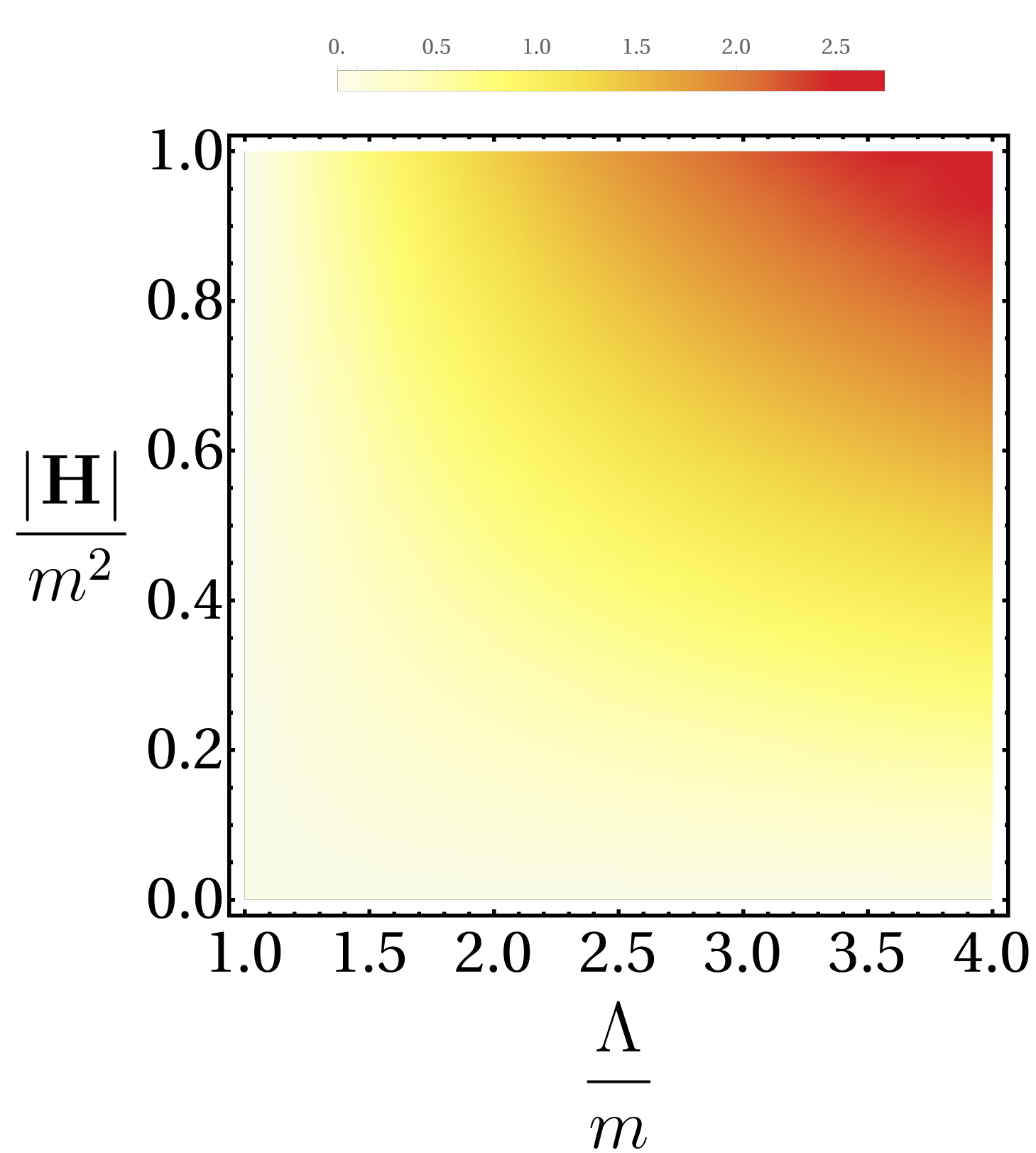}
  \caption{Heat-map of $2^3 3 \pi^2 \abs{\boldsymbol{\Delta}_1}$ as a function of the absolute value of the chromo-magnetic field and energy scale $\Lambda$.}
  \label{BHFig3m}
\end{minipage}
\end{figure}

In Figs.
\ref{BHFig} and \ref{BHFig3m}
the quantity
$|\Delta_1|$ (with $\mathbf{E}=0$)
 respectively
for a small range of the cutoff 
around $\Lambda \sim m$
and  for larger values of the cutoff.
In the first case, for lower values of the cutoff, 
negative values for $|\Delta_1|$
contributes for  the c-magnetic diamagnetic
 character of the vacuum.
For larger values of the curtoff the overall 
correction $|\Delta_1|$ contributes for c-paramagnetic vacuum.
However a complete analysis may require a numerical compuation 
of the full (non-expanded) determinant (or higher order
 contributions) and it 
requires
further gluon and ghost determinants.

 \subsection{ Some conditions for 
 zero imaginary quantities in the Abelian limit}
\label{AbelianLimit}

In this section, we rewrite some equations in the Abelian limit for the EH electromagnetic case.
This makes explicit the corresponding conditions 
for real electromagnetic observables.

The trace of the energy-momentum tensor \eqref{TRACEEMT}
in the Abelian limit can be written as:
\begin{align}\label{nonA-TRACEEMT}
    T^\mu_\mu =& - 8 \frac{\lambda_2}{m^2} (j^\mu)^2 
    + 16 \frac{(\lambda_3+ \lambda_6)}{m^4} (F^2)^2
   - 4 \frac{  \lambda_8  }{m^4}  (\partial_\mu j^\nu) F^{\mu\alpha} F_{\alpha \nu} 
    - 6  \frac{\lambda_{10}}{m^4}(\partial_\mu j_\nu)^2
    + m \Bar{\psi} \psi.
\end{align}
There still is an imaginary part in the fermion-photon field interaction for time-dependent and/or spatial non-homogeneous fermion density and current. 
The second term, $(F^2)^2$, is  exclusively due to the photon self interaction.
All the others, except the last one that is the classical mass term from the Lagrangian,
represent  effective fermion  self interactions and photon-fermion higher order
interaction.
The correspondence  to the non-Abelian cases discussed above is very clear for 
each term.

 The Poyinting vector \eqref{vector-Poyinting} in the Abelian limit becomes:
\begin{eqnarray}    
    \nonumber
    \mathbf{S} &\equiv& \sum_{i=1}^{3}T^{i0} \mathbf{e}_i = \left( 1 + \frac{g^2}{24\pi^2} \ln \frac{\Lambda^2}{m^2} \right) \mathbf{E}_a \times \mathbf{H}_a 
    +64 \frac{\lambda_3}{m^4}   \left( \mathbf{E} \times \mathbf{H} \right) \left( \mathbf{E}^2 - \mathbf{H}^2\right)  
    + 32 \frac{\lambda_6}{m^4}   \mathbf{E} \times \mathbf{H}
    \left( \mathbf{H}^2- \mathbf{E}_c^2 \right) + 
    \\
    \nonumber
    &+& 
   4 \frac{  \lambda_8 }{m^4}
   \left( 
    \left\{ (\nabla \rho) - (\partial_t \mathbf{j}) \right\} \mathbf{E}^2 + \left\{  (\nabla j^j) - 
    (\partial^j \mathbf{j}) \right\} \left( \mathbf{H} \times \mathbf{E} \right)^{j} 
    +\left\{ \left((\nabla \rho) - (\partial_t \mathbf{j}) \right\} \cdot \mathbf{E} \right) 
    \mathbf{E} + \left\{ (\partial^j \mathbf{j}) - (\nabla j^j) \right\} \times \mathbf{H} E^j \right) 
\\\nonumber
    &+&
    2 \frac{\lambda_{10}}{m^4} \left( (\nabla \rho)_a (\partial_t \rho)_a - (\nabla j^j)_a (\partial_t j^j)_a \right)
    + \frac{i}{2} \left( \Bar{\psi} \boldsymbol{\gamma} 
     \partial_t\psi - (\partial_t \Bar{\psi}) \boldsymbol{\gamma} \psi +  \Bar{\psi} \gamma^0 \nabla \psi - (\nabla \Bar{\psi}) \gamma^0 \psi \right).
\end{eqnarray}
There are several imaginary terms for the 
coupling $\lambda_8$.
Emission and absorption of electromagnetic waves correspond to at least two terms with imaginary coefficients.

 The induced fields $\mathbf{D}$ and 
 $\mathbf{B}$ 
 can be written as
\begin{align}\nonumber
\mathbf{D} =& \left( 1 + \frac{ g^2}{24 \pi^2} \ln{\frac{\Lambda^2}{m^2}} \right) \mathbf{E} + \frac{32 (2\lambda_3 + \lambda_6 )}{m^4} \mathbf{E} (\mathbf{E}^2) - \frac{4\lambda_8}{m^4} \left( \left( \nabla j^i \right) E^i + \left( \mathbf{E} \cdot \nabla \right) \mathbf{j} - 2 \left(\partial_t \rho \right) \mathbf{E} \right)
   ,
    \\
    \mathbf{B} =& \left( 1 + \frac{g^2}{24 \pi^2} \ln{\frac{\Lambda^2}{m^2}} \right) \mathbf{H}
    - \frac{32
    (2\lambda_3+ \lambda_6)}{m^4}  \mathbf{H}   (\mathbf{H}^2) 
    + \frac{4 \lambda_8}{m^4}   \left( 
    \left( \nabla  j^i \right) H^i + \left(\mathbf{H} \cdot \nabla  \right) \mathbf{j} - 2 ( \nabla  \cdot \mathbf{j} ) 
    \mathbf{H} \right) 
      .
\end{align}
These fields will be real for the following situations:
\begin{eqnarray}
\Im \mathbf{D} = 0, 
\;\;\;\;\;
 \left( \nabla  j^i \right) H^i + \left(\mathbf{H} \cdot \nabla  \right) \mathbf{j} - 2 ( \nabla  \cdot \mathbf{j} ) 
    \mathbf{H}  &=& 0,
\nonumber
\\
\Im \mathbf{B} = 0, 
\;\;\;\;\;\;\;
 \left( \nabla j^i \right) E^i + \left( \mathbf{E} \cdot \nabla \right) \mathbf{j} - 2 \left(\partial_t \rho \right) \mathbf{E} 
 &=& 0.
\end{eqnarray} 
These conditions indicate that imaginary terms for 
the permittivity  (permeability)
are due to external magnetic (electric) fields and fermion currents or time dependent density.

\section{ Summary}
\label{sec:summary}

We have presented a large quark mass expansion of the quark determinant in the presence of a gluon background field
by including quark current sources,
%along the   derivative expansion proposed by Novikov {\it et al}
for the local limit of the effective action.
%\cite{Novikov1984}.
As a result, we exhibitted not only contributions for 
 the gluon self-interaction terms, known as Euler-Heisenberg-type QCD action  computed previously by several authors, 
but also different effective quark-gluon interactions and quark self  interactions.
Besides the  self interactions also 
present in the Abelian effective action,  
the QCD effective action presents several interactions with strict non-Abelian character.
Interestingly, some of them 
disappear when written in terms of chromo-electromagnetic fields corresponding to 
{\it trivial } (topological) constraints, 
mainly in the   gluon sector.
In this expansion of the quark determinant,   
all the terms allowed by symmetry up to the order of mass$^8$
were included and results are valid for 
non-homongeneous gauge fields and quark currents. 
The  Abelian limit, for
the electromagnetic effective action, was also obtained.
It  contains
  several types of  photon-fermion current effective  interactions
and few fermion self-interactions.
The validity of the large quark mass expansion was analyzed by means of equations
\eqref{convergence1} and \eqref{convergence2}.
Although the expansion is usually recognized to be valid for 
the heavy quark sector, it can be seen that it may be valid for the regime 
of the constituent quark model in which quark masses receive 
large contributions from 
mass generation mechanisms.

For the QCD-EH action, 
most of the different higher-order 
quark-gluon interactions
are  dependent on  space and time variations of the 
quark density and  current.
As exceptions to these, there 
are the corrections to the quark-quark (or quark-antiquark interaction)
terms
$\lambda_2$ and  $ \lambda_{10}$.
Several properties of the resulting effective action, and its Abelian limit,
were calculated such as 
the energy-momentum tensor and its trace, 
the Poyinting vector, the chromo-permeability and the chromo-permittivity.
Several of  the interactions contribute for the trace anomaly, as seen in section \eqref{traceanomaly}.
As a consequence, effective models that consider effective quark self interactions
might be considered as to have trace anomaly incorporated in their effective  couplings.
This becomes still clearer when writing the Abelian limit in section \eqref{AbelianLimit}
which contains some of the (more intrincated) terms of the non-Abelian action. 
The higher order (effective) quark-gluon interactions  terms have different structures,
some of them 
reminding chromo-electromagnetic wave
coupling
(emission or absorption)
to varying quark current and density
or higher order chromo-electric field coupling to the time-dependent
quark density.
This analysis is clearer in 
 the Abelian limit case in section \eqref{sec:abelian}.

A  general condition for the reality of the EH-QCD
effective action can be written as:
\begin{eqnarray}
\nonumber
 0  &=& d_{abc}  \left[ \left( \mathcal{D}_t \rho_a \right) \mathbf{E}_b \cdot \mathbf{E}_c - \left( \mathcal{D}_t \mathbf{j}_a \right) \cdot \mathbf{E}_b \times \mathbf{H}_c -
   \left( \mathbf{E}_b \cdot \boldsymbol{\mathcal{D}} \;  \mathbf{j}_a 
    \; \cdot \right) \mathbf{E}_c  + \left( \boldsymbol{\mathcal{D}} \rho_a \right) \cdot \left( \mathbf{H}_b \times \mathbf{E}_c \right) \right.
    \\
    \nonumber
    &+& \left. \left( \boldsymbol{\mathcal{D}} \cdot \mathbf{j}_a \right) \mathbf{H}_b \cdot \mathbf{H}_c - \left( \mathbf{H}_c \; \cdot \boldsymbol{\mathcal{D}} \mathbf{j}_a \; \cdot \right) \mathbf{H}_b \right],
    \end{eqnarray}
    for which one may have
    particular cases such as
$ \mathbf{H}_c = 0 $ or $ \mathbf{E}_c=0$, with corresponding conditions in Eq. 
\eqref{cond-im=0}.
%The quark-gluon higher-order interactions
%of the effective action
%\eqref{Seffresult} 
%can also be expected to be useful for 
%QCD low-energy effective models.
%The trace of the energy-momentum tensor was also presented, 
% containing the different types of effects from the gluon sector and quark-gluon interaction sector.
The 
chromo-magnetic induction $B$
 and chromo-electric field $D$
 were   calculated in the presence of matter density and currents, exhibiting strong non-linear behavior.
Conditions for the reality of these
observables, and their Abelian limits, were also established.
Note that the conditions for the reality of the effective action, for the energy momentum tensor and its trace,
and the conditions for real induction fields $B$ and $D$ are not the same.
This gives rise to the possibility of instabilities happen in different sectors of the 
resulting effective action (i.e. c-electric or c-magnetic or overall energy and so on)
at different points.
These possibilities are intended to be investigated in detail in another work.
The contributions from the gluon and ghost determinants were left outside the scope of this work.

Numerical estimations for some quantities such as 
energy density  and c-electric permittivity and c-magnetic
permeability have been provided for two different 
contexts: relatively small cutoff ($\Lambda \sim m$),
seemingly for the case of a somewhat heavy quark,
and relatively larger cutoff ($\Lambda \sim 3 m$),
seemingly for the case of a somewhat light  quark.
This variation of the cutoff might manifest in 
the renormalized equations due to the 
renormalization energy scale that may be 
seen to vary according to 
the cutoff.
Numerical results are somewhat different for each of these two
cases.
In particular for the small cutoff context,
the resulting EFT contributes for 
 more accentuated
chromo-(dia)magnetic behavior
with more negative energy density.

Concerning the Abelian limit, 
there are imaginary terms in the fermion-photon interaction sector
that can be associated not only with pair production but, seemingly, also with 
absorption and emission of electromagnetic waves.
This can be noted from the
Poynting vector  
coupling to time-dependent or spatial varying quark densities and currents. 
These structures and correspoding effects 
also manifest , at least partially, in the non-Abelian case, being that the Poyinting vector  shows a considerably more intricate structure.
The condition for the real 
Abelian electromagnetic EH action with sources is given by
\begin{eqnarray}
0 = \left( - (\mathbf{E}  (\mathbf{E} \cdot \nabla) + \mathbf{H} (\mathbf{H} \cdot \nabla))  \cdot \mathbf{j} + \frac{\partial \rho}{\partial t} \mathbf{E}^2 + \mathbf{H}^2 \nabla \cdot \mathbf{j}   + \mathbf{H} \times \mathbf{E} \cdot \left( \frac{\partial \mathbf{j}}{\partial t} + \nabla \rho \right) \right).
\nonumber
\end{eqnarray}
There may exist several possible configurations that   satisfy this condition,
either in the electric or 
in the magnetic sectors, or in mixed configurations.
Therefore, 
it is possible to envisage 
fine-tuning time-dependent and spatial-non-homogeneous 
configurations that assure real-valued action and corresponding dynamics. 
By 
switching off these conditions,
the imaginary terms can contribute to destabilizing the system. 
Or vice versa:
out of a system with these terms that drive pair creation or absorption/emission of electromagnetic field configurations, such as light, one 
can search for configurations that drive these imaginary terms to zero, by fine-tuning of the time and spatial behavior of matter density and current for a particular wave frequency or wavelength.
Notwithstanding, other (non)perturbative contributions 
might appear to a   complete
Low-energy effective action and a calculation for non-large fermion mass
 can bring higher order corrections to these terms
and modify the coefficient $\lambda_8$.
However the shape of the interaction terms are defined by symmetry \cite{Weinberg1995}.

\subsection*{ Acknowledgements}

The authors thank a short conversation with A.J. Mizher
and A. Cucchieri.
F.L.B. is member of INCT-FNA 
 464898/2014-5.
I. de M. F. acknowledges  financial support  from CAPES, 
and F. L. B. acknowledges  partial support from
CNPq-312750/2021-8 and CNPq-407162/2023-2.

\appendix
\section{ Derivative expansion and
Identities}
\label{app:A}

In this Appendix, several 
important manipulations and identities needed to calculate the expansion of the effective action are presented.

\subsection{ 
Derivative expansion 
}

The gluon field and
its consequences encoded in
the covariant derivative
in $D$ must be developed.
Terms of the following form, 
for a function F(G,J) of $G$ and $J$,
will arise:
\begin{eqnarray}
Tr \left[
D^n \; F(G,J)
\right].
\end{eqnarray}
If the operators in $F(G,J)$
have an overall dimension smaller
than [mass]$^8$, this expression must be developed (in the present case, expanded)
to produce terms of the maximal order [mass]$^8$ with the 
gluon fields in the operator $D$.

For that, 
consider
a regularized generic basic term $S$ for which, being finite,
a momentum shift $p_\mu \to 
p_\mu + q_\mu$ can be done in the integrand.
It can be written that:
\begin{eqnarray}
S &=& 
Tr \; \left[ 
( D - D_0 ) F(G,J)
\right]
\to 
Tr \; \left[ 
\left( \frac{1}{
(\Pi - q)^2 - m^2
} - 
\frac{1}{(p-q)^2 - m^2}) \right) F(G,J)
\right],
\end{eqnarray}

The derivative expansion, as proposed in
\cite{Mosel2004}, can be implemented as it follows:
\begin{eqnarray}
T &=& - \frac{\sigma \cdot F}{2}, \;\;\;\;\;
D^{-1} = [ (\partial^\mu - g A^\mu ) (\partial_\mu - g A_\mu)  - M^2]
\\
&& \sum_{n=0}^{\infty}
 ( D T )^n ,
\\
T D &=& D T + D [ D^{-1}, T ] D
=
D T + D^2 [ D^{-1} , T ] + D^2 [ D^{-1} , [ D^{-1} , T ] D + ...
\end{eqnarray}

\subsection{Identities}
\label{Apendice_Identidades}

The required identities for implementing the procedure up to [mass]$^8$ are given below:
\begin{align}\label{Ident0}
    \Tr{D^3 F(G,J)} =& -\frac{i}{2^5 \pi^2 (m^2)} \Tr_{C,F}{F(G,J)} + \frac{ig^2}{2^6 3 \pi^2 (m^2)^3} \Tr_{C,F}{G^2 F(G,J)};\\ \nonumber
    \Tr{D^4 \Pi^\alpha F(G,J)} =& \frac{ig}{2^7 \cdot  3 \cdot \pi^2 \cdot (m^2)^3} \Tr_{C,F}{J^\alpha F(G,J)} \\\nonumber
    & + \frac{g^2}{2^6 \cdot 3^2 \cdot 5 \cdot \pi^2 \cdot (m^2)^4} \Tr_{C,F} \left\{  \left( -6g \d^\alpha G^2 + 66g G^{\alpha \rho} J_\rho -10g (\d^\mu G^{\alpha \rho})G_{\rho \mu}  \right. \right. \\ \nonumber
    & - \frac{9i}{2} \d^2 J^\alpha + \frac{3i}{2} \d^\rho \d^\alpha J_\rho + 6i \d_\mu \d^2 G^{\alpha \mu} + 2g G^{\rho \mu} \d^\alpha G_{\rho \mu} \\\nonumber
    &  -3g G_{\phi \mu} \d^\phi G^{\alpha \mu} -2g \{ J_\rho, G^{\rho \alpha}  \} -2g \{ G_{\phi \rho}, \d^\phi G^{\rho \alpha}\} \\\nonumber
    &  + 2g(\d^\alpha G^{\mu \rho})G_{\rho \mu} + 2g (\d^\mu G^{\alpha \rho}) G_{\rho \mu} - 45g J_\rho G^{\rho \alpha} \\\label{Ident1}
    &\left. \left. -3g J_\rho G^{\rho \alpha} \right) F(G,J) \right\} \\ \nonumber
    \Tr{D^5\{ \Pi^\alpha, \Pi^\beta \} F(G,J)} =& \frac{i}{2^7 \cdot 3 \cdot \pi^2 \cdot (m^2)^4} \Tr_{C,F}{g^{\alpha \beta} F(G,J)} \\\nonumber
    & - \frac{ig^2}{2^8 \cdot 3 \cdot \pi^2 \cdot (m^2)^4} \Tr_{C,F}{ \left( G^2 g^{\alpha \beta} + 2 \{ G^{\alpha \mu}, G_\mu ^\beta \} \right) F(G,J)}\\\label{Ident2}
    &- \frac{g}{2^7 \cdot 3 \cdot \pi^2 (m^2)^4} \Tr_{C,F}\left\{  \left( \mathcal{D}^\alpha J^\beta + \mathcal{D}^\beta J^\alpha \right) F(G,J) \right\};\\\nonumber
    \Tr{D^6 \Pi^\alpha \Pi^\beta \Pi^\mu F(G,J)} =& - \frac{11ig}{2^{10} \cdot 3^2 \cdot 5 \pi^2 (m^2)^4} \Tr_{C,F}{J^\alpha g^{\beta \mu} + J^{\beta} g^{\mu \alpha} + J^\mu g^{\alpha \beta}} \\\label{Ident3}
    &- \frac{ig}{2^6 \cdot 3 \cdot 5 \pi^2 (m^2)^4} \Tr_{C,F}{\mathcal{D}^\mu G^{\beta \alpha} + \mathcal{D}^\beta G^{\mu \alpha}};\\\nonumber
    \Tr{D^7 \Pi^\alpha \Pi^\beta \Pi^\mu \Pi^\nu F(G,J)} =& \frac{-i}{2^9 \cdot 3^2 \cdot 5 \cdot \pi^2 \cdot (m^2)^3} \Tr_{C,F}{\left( g^{\alpha \beta} g^{\mu \nu} + g^{\alpha \mu} g^{\beta \nu} + g^{\alpha \nu} g^{\beta \mu} \right)F(G,J)}\\\nonumber
    &- \frac{g}{2^9 \cdot 3 \cdot 5 \cdot \pi^2 \cdot (m^2)^4} \Tr_{C,F}\left\{\left(  g^{\alpha \beta} G^{\mu \nu} + g^{\alpha \mu} G^{\beta \nu} \right. \right.\\\label{Ident4}
    & \left. \left.  + g^{\alpha \nu} G^{\beta \mu} + g^{\beta \mu} G^{\alpha \nu} + g^{\beta \nu} G^{\alpha \mu} + g^{\mu \nu} G^{\alpha \beta} \right)F(G,J) \right\}
\end{align}

\subsection{Commutators}
\label{Apendice_Comutadores}

Here we present the needed commutators up to [mass]$^6$. They are a generalization of the source case to the ones found in \cite{Novikov1984}:
\begin{align}
    [\Pi^2, \Pi_\mu] =& 2ig \Pi^\alpha G_{\alpha \mu} + gJ_\mu;\\\nonumber
    [\Pi^2, [\Pi^2, \Pi_\mu]] =& -4g \Pi^\alpha \Pi^\beta \mathcal{D}_\beta G_{\alpha \mu} + \Pi^{\alpha} \left( -4g^2 G_{\alpha}^{\;\; \beta} G_{\beta \mu} + 2ig \mathcal{D}^2 G_{\alpha \mu} + 2ig \mathcal{D}_\alpha J_\mu \right)\\\nonumber
    &+2ig^2 J^\alpha G_{\alpha \mu} + g \mathcal{D}^2 J_\mu;\\\nonumber
    [\Pi^2, [\Pi^2, [\Pi^2, \Pi_\mu]]] =& -8ig \Pi^\alpha \Pi^\beta \Pi^\gamma \mathcal{D}_\gamma \mathcal{D}_\beta G_{\alpha \mu} + \Pi^\alpha \Pi^\beta \left( -8ig^2 G_\alpha^{\;\; \nu} \mathcal{D}_\beta G_{\nu \mu} - 8ig^2 G_{\beta}^{\;\; \nu} \mathcal{D}_\nu G_{\alpha \mu} \right.\\\nonumber
    &\left. - 4g \mathcal{D}^2 \mathcal{D}_\beta G_{\alpha \mu}  -4g^2 \mathcal{D}_\beta (G_{\alpha}^{\;\; \nu} G_{\nu \mu}) - 4g \mathcal{D}_\beta \mathcal{D}^2 G_{\alpha \mu} - 4g \mathcal{D}_\beta \mathcal{D}_\alpha J_\mu) \right) \\\nonumber
    & + \Pi^\alpha \left( -8g^2 (\mathcal{D}_\gamma G_{\alpha}^{\;\; \beta}) \mathcal{D}^\gamma G_{\beta \mu}  \right. - 4g^2 J^\beta \mathcal{D}_\alpha G_{\beta \mu} - 4g^2 J^\beta \mathcal{D}_\beta G_{\alpha \mu} - 8ig^3 G_{\alpha}^{\;\; \beta} G_{\beta}^{\;\; \gamma} G_{\gamma \mu} \\\nonumber
    & \left.  - 4g^2 G_{\alpha}^{\;\; \beta} \mathcal{D}^2 G_{\beta \mu} - 4g^2 G_{\alpha \beta} \mathcal{D}^\beta J_\mu - 4g^2 \mathcal{D}^2 (G_{\alpha}^{\;\; \beta} G_{\beta \mu}) + 2ig \mathcal{D}^4 G_{\alpha \mu} + 2ig \mathcal{D}^2 \mathcal{D}_\alpha J_\mu \right.\\
    &\left.+ 2ig \mathcal{D}_\alpha \mathcal{D}^2 J_\mu - 4g^2 \mathcal{D}^\alpha \mathcal{D}^2(J^\beta G_{\beta \mu}) \right)\\\nonumber
    [\Pi^2, [\Pi^2, [\Pi^2, [\Pi^2, \Pi_\mu]]]] =& 16 g \Pi^{\alpha} \Pi^{\beta} \Pi^{\gamma} \Pi^{\phi} \mathcal{D}_\phi \mathcal{D}_{\gamma} \mathcal{D}_{\beta} G_{\alpha \mu} + \Pi^{\alpha} \Pi^{\beta} \Pi^{\gamma} \left( 16g^2 G^{\alpha \nu} \mathcal{D}_{\gamma} \mathcal{D}_{\beta} G_{\nu \mu} + 16g^2 G^{\beta \nu} \mathcal{D}_\gamma \mathcal{D}_\nu G_{\alpha \mu} \right.\\\nonumber
    & + 16g^2 G_{\gamma \nu} \mathcal{D}^\nu \mathcal{D}_\beta G_{\alpha \mu} - 8ig \mathcal{D}^2 \mathcal{D}_\gamma \mathcal{D}_\beta G_{\alpha \mu} + 16g^2 \mathcal{D}_\gamma \left( G_{\alpha}^{\;\; \nu} \mathcal{D}_\beta G_{\nu \mu} \right)  \\\nonumber
    & + 16g^2 \mathcal{D}_\gamma \left( G_{\beta \nu} \mathcal{D}^\nu G_{\alpha \mu} \right) -8ig \mathcal{D}_\gamma \mathcal{D}^2 \mathcal{D}_\beta G_{\alpha \mu} - 8ig^2 \mathcal{D}_\gamma \mathcal{D}_\beta \left( G_{\alpha}^{\;\; \nu} G_{\nu \mu}\right) \\
    & \left. - 8ig \mathcal{D}_\gamma \mathcal{D}_\beta \mathcal{D}^2 G_{\alpha \mu} - 8ig \mathcal{D}_\gamma \mathcal{D}_\beta \mathcal{D}_\alpha J_\mu \right)\\
    [\Pi^2, [\Pi^2, [\Pi^2, [\Pi^2, [\Pi^2, \Pi_\mu]]]]] =& 32ig \Pi^\alpha \Pi^\beta \Pi^\gamma \Pi^\phi \Pi^\lambda \mathcal{D}_\lambda \mathcal{D}_\phi \mathcal{D}_\gamma \mathcal{D}_\beta G_{\alpha \mu}.
\end{align}

\subsection{Structure of terms from the expansion}

The resulting low-energy effective field theory can be written in terms of the following quantities:
$\mathcal{D}_\mu$, 
$G_{\mu\nu}$ and 
$J_\mu$.
The expansion was truncated 
in the order $M^8$,
with the following conditions 
\begin{eqnarray}
2k+l+3n \leq 8, 
\;\;\;\;\; 2k+l+n = 2,4,6...
\end{eqnarray}.
Accordingly, the following terms can be written in the operator expansion:
\begin{table}[H]
\hspace{-0.5cm}
\centering
\begin{tabular}{|c|c|c|c|c|c|c|c}
\cline{1-7}
  & k & l     & n &  2k+l+3n & 2k+l+n & Terms  &  \\ \cline{1-7}
a & 4 & 0     & 0 & 8       & 8      & EH     &  \\ \cline{1-7}
b & 3 & 0     & 0 & 6       & 6      & EH     &  \\ \cline{1-7}
c & 3 & 2     & 0 & 8       & 8      & $\left( \mathcal{D}_\mu \mathcal{D}_\nu G_{\alpha \beta} \right) G^{\mu \nu} G^{\alpha \beta}$ &  \\ \cline{1-7}
d & 2 & 0     & 0 & 4       & 4      & EH     &  \\ \cline{1-7}
e & 2 & 1     & 1 & 8       & 6      &   $G^{\mu\nu}G_{\mu \alpha} \mathcal{D}^\alpha J_\nu$, $\left(\mathcal{D}^\alpha G^{\mu\nu} \right) G_{\mu \alpha} J_\nu$     &  \\ \cline{1-7}
f & 2 & 2,4   & 0 & 6,8       & 6,8      & $\left( \mathcal{D}_\mu \mathcal{D}_\nu G^{\mu \alpha} \right) G_{\alpha}^\nu$, $\left( \mathcal{D}_\mu \mathcal{D}_\nu \mathcal{D}_\alpha \mathcal{D}_\beta G^{\mu \nu} \right) G^{\alpha \beta}$ &  \\ \cline{1-7}
g & 1 & 0     & 0 & 2       & 2      & EH = 0     &  \\ \cline{1-7}
h & 1 & 1     & 1 & 6       & 4      &    $G_{\mu \nu} \mathcal{D}^\mu J^\nu$    &  \\ \cline{1-7}
i & 1 & 3     & 1 & 8       & 4      &    $G_{\mu \nu} \mathcal{D}^\alpha \mathcal{D}^\mu \mathcal{D}^\nu J_\alpha$    &  \\ \cline{1-7}
j & 1 & 2,4,6 & 0 & 4,6,8   & 4,6,8  & $\mathcal{D}_\mu \mathcal{D}_\nu G^{\mu \nu}$, $\mathcal{D}_\mu \mathcal{D}^\alpha \mathcal{D}_\nu \mathcal{D}_\alpha G^{\mu \nu}$, $ \mathcal{D}_\beta \mathcal{D}^\beta \mathcal{D}_\mu \mathcal{D}_\alpha \mathcal{D}_\nu \mathcal{D}^\alpha G^{\mu \nu}$&  \\ \cline{1-7}
k & 1 & 0     & 2 & 8       & 4      &    $G_{\mu \nu} J^\mu J^\nu$    &  \\ \cline{1-7}
l & 0 & 0     & 2 & 6       & 2      &     $J_\mu J^\mu$   &  \\ \cline{1-7}
m & 0 & 2     & 2 & 8       & 4      &    $\mathcal{D}_\mu \mathcal{D}_\nu J^\mu J^\nu$    &  \\ \cline{1-7}
n & 0 & 1     & 1 & 4       & 2      &    $\mathcal{D}_\mu J^\mu = 0$    &  \\ \cline{1-7}
o & 0 & 3     & 1 & 7       & 4      &    $\mathcal{D}_\mu \mathcal{D}_\nu \mathcal{D}^\mu J^\nu$    &  \\ \cline{1-7}
\end{tabular}
\caption{
Terms that possibly arise in expansion up to the order [mass]$^8$. Terms with ``EH'' stand for those arising in the Euler-Heisenberg action. }
\label{Tab1}
\end{table}

Changes in the index and positions of the fields are possible by considering the operations discussed in the text.
Several of these terms are not linearly independent since
considering the identities and operations discussed in the text terms presented in one line may be written as combinations of terms of the other lines.

\section{ Explicit calculation
}

\label{app:coefficients}

To clarify how the computations of each coefficient in table \ref{tab:couplingsSeff} are done from Eq. \eqref{ScalarQCDS}, we demonstrate how to get the contribution from the first commutator $d^3 \Gamma_{q}^{(1)}/d (m^2)^3 \equiv - (i/m^2) \Tr \left( D^4 [\Pi^2, \Pi_\mu] \Pi^\mu \right) $ to each effective coupling $g_i$ or $\lambda_i$, for sQCD or QCD respectivelly. Using the commutators in the Appendix \ref{Apendice_Comutadores}, we may write it as:
\begin{align}
    \frac{d^3 \Gamma_{q}^{(1)}}{d (m^2)^3} =& - \frac{i}{m^2} \Tr{ \left( D^4 \left\{ 2ig \Pi^\alpha G_{\alpha \mu} + gJ_\mu \right\}  \Pi^{\mu} \right) }.
\end{align}

To integrate in momentum, we must separate the above expression into local in $x$ and $p$ parts. This is done by simply using the commutator of the variables, \textit{i.e.} $F(G,J) \Pi^\mu = \Pi^\mu F(G,J) + [F(G,J), \Pi^\mu] = \Pi^\mu F(G,J) -i \mathcal{D}^{\mu} F(G,J) $, where $F(G,J)$ is a general function of the gluon and source fields. Then:
\begin{align}\nonumber
    \frac{d^3 \Gamma_{q}^{(1)}}{d (m^2)^3} =& - \frac{i}{m^2} \Tr{ \left( D^4 \left\{ 2ig \Pi^\alpha \Pi^\mu G_{\alpha \mu} + 2g \Pi^\alpha \mathcal{D}^\mu G_{\alpha \mu} + g \Pi^\mu J_\mu -ig \mathcal{D}^\mu J_{\mu} \right\}  \right) }\\\nonumber
    =& - \frac{i}{m^2} \Tr{ \left( D^4 \left\{ ig [\Pi^\alpha, \Pi^\mu] G_{\alpha \mu} + ig \{\Pi^\alpha, \Pi^\mu\} G_{\alpha \mu} - 2g \Pi^\alpha J_{\alpha} + g \Pi^\mu J_\mu  \right\}  \right) }\\\label{FirstCommutator}
    =& - \frac{i}{m^2} \Tr{ \left( D^4 \left\{ -g^2 G^{\alpha \mu} G_{\alpha \mu} + ig \{\Pi^\alpha, \Pi^\mu\} G_{\alpha \mu} - g \Pi^\alpha J_{\alpha} \right\}  \right) },
\end{align}
where we have used the equation of motion of the gluon field, the continuity equation and the definition of the gluon field, given respectively by:
\begin{align} \label{EOM}
    \mathcal{D}_\mu G^{\mu \alpha} =& J^\alpha;\\
    \mathcal{D}_\mu J^\mu =& 0:\\
    [\Pi^\mu, \Pi^\alpha] =& ig G^{\mu \alpha}.
\end{align}

Notice that the second term on the left-hand side of E. \eqref{FirstCommutator} is zero due to the antisymmetric nature of $G_{\alpha \mu}$. Now, since we want to go up to order [mass]$^8$, we must use the integral expressions for:
\begin{itemize}
    \item $\Tr D^3 F(G,J)$ up to [mass]$^4$;
    \item $\Tr D^3 \Pi^{\mu} F_{\mu}(G,J)$ up to [mass]$^5$,
\end{itemize}
which are given in Appendix \ref{Apendice_Identidades}. Then, Eq. \eqref{FirstCommutator} becomes:
\begin{align}\nonumber
    \frac{d^3 \Gamma_{q}^{(1)}}{d (m^2)^3} =& - \frac{i}{m^2}  \Tr_{C,F} \left( -\left\{ \frac{i}{2^6 3 \pi^2 (m^2)^2} - \frac{ig^2}{2^6 3 \pi^2 (m^2)^3 } G^2 \right\}g^2 G^2 \right. \\\nonumber
    & \left.- \bigg\{ \frac{ig}{2^7 3 \pi^2 (m^2)^3} J^\alpha + \frac{g^2}{2^6 3^2 5 \pi^2 (m^2)^4} \left( -6g \d^\alpha G^2 + 66g G^{\alpha \rho} J_\rho -10g (\d^\mu G^{\alpha \rho})G_{\rho \mu} - \frac{9i}{2} \d^2 J^\alpha  \right. \right. \\ \nonumber
    &  + \frac{3i}{2} \d^\rho \d^\alpha J_\rho + 6i \d_\mu \d^2 G^{\alpha \mu} + 2g G^{\rho \mu} \d^\alpha G_{\rho \mu}  -3g G_{\phi \mu} \d^\phi G^{\alpha \mu} -2g \{ J_\rho, G^{\rho \alpha}  \} -2g \{ G_{\phi \rho}, \d^\phi G^{\rho \alpha}\} \\\nonumber
    &  + 2g(\d^\alpha G^{\mu \rho})G_{\rho \mu} + 2g (\d^\mu G^{\alpha \rho}) G_{\rho \mu} - 45g J_\rho G^{\rho \alpha} \left. \left. -3g J_\rho G^{\rho \alpha} \right. \bigg) \bigg\} gJ_{\alpha}  \right).
\end{align}

This expression contributes to all effective couplings in table \ref{tab:couplingsSeff}, using integration by parts, the continuity equation, the definition of the gluon field and Jacobi's identity.

% \Tr_{C,F}{J^\alpha F(G,J)} \\\nonumber
%    & + \frac{g^2}{2^6 \cdot 3^2 \cdot 5 \cdot \pi^2 \cdot (m^2)^4} \Tr_{C,F} \left\{  \left( -6g \d^\alpha G^2 + 66g G^{\alpha \rho} J_\rho -10g (\d^\mu G^{\alpha \rho})G_{\rho \mu}  \right. \right. \\ \nonumber
%    & - \frac{9i}{2} \d^2 J^\alpha + \frac{3i}{2} \d^\rho \d^\alpha J_\rho + 6i \d_\mu \d^2 G^{\alpha \mu} + 2g G^{\rho \mu} \d^\alpha G_{\rho \mu} \\\nonumber
%    &  -3g G_{\phi \mu} \d^\phi G^{\alpha \mu} -2g \{ J_\rho, G^{\rho \alpha}  \} -2g \{ G_{\phi \rho}, \d^\phi G^{\rho \alpha}\} \\\nonumber
%    &  + 2g(\d^\alpha G^{\mu \rho})G_{\rho \mu} + 2g (\d^\mu G^{\alpha \rho}) G_{\rho \mu} - 45g J_\rho G^{\rho \alpha} \\\label{Ident1}
%    &\left. \left. -3g J_\rho G^{\rho \alpha} \right) F(G,J) \right\}

\section{ Corrections to Yang-Mills
and Maxwell equations}

\label{app:YangMillseqs}

By using the Euler-Lagrange   equations,   the corrected Yang-Mills equations
for the effective action \eqref{Seffresult} will be given by:
% \begin{align}\nonumber
%         & - \frac{\lambda^2}{\lambda_R^2} \mathcal{D}_\mu G^{\mu \nu}_a - \frac{\lambda_1}{m^2} \mathcal{D}^\mu \left( f_{abc} G_{\mu \rho b} G^{\rho \nu}_c \right) + 8 \left[ \frac{\lambda_3}{m^4} k_{abcd} + \frac{\lambda_5}{m^4} f_{ace} f_{dbe} \right] \mathcal{D}_\mu \left( G^{\mu \nu}_b G_{\alpha \beta c} G^{\alpha \beta}_d \right)\\\nonumber
%         & + 4 \left( \frac{\lambda_4}{m^4}f_{abe}f_{cde} + \frac{\lambda_6}{m^4} k_{abcd} \right) \mathcal{D}_\mu \left( G^{\lambda \nu}_b G^{\mu \beta}_{c} G_{\beta \nu d} - G^{\mu \nu}_b G^{\lambda \beta}_{c} G_{\beta \nu d} \right)\\\nonumber
%         & + \left( -\frac{\lambda_7}{m^4} f_{abc} + \frac{\lambda_8}{m^4} d_{abc} \right) \mathcal{D}_\mu \left( \left( \mathcal{D}^\mu J_{\alpha b} \right) G^{\nu \alpha}_c - \left( \mathcal{D}^\nu J_{ \alpha b } \right) G^{\mu \alpha}_c + \left( \mathcal{D}_\alpha J_{ \nu c } \right) G^{\alpha \mu}_b - \left( \mathcal{D}^\alpha J_{ \mu c } \right) G^{\alpha \nu}_b \right)\\\nonumber
%         & +2 \frac{\lambda_9}{m^4} f_{abc} \mathcal{D}_\mu \left( J^{\mu}_b J^\nu_c \right) = - J^\nu_a + 2\frac{\lambda_{10}}{m^4} gf_{abc} J^\mu_b \left( \mathcal{D}^\nu J_{\mu c} \right) + f_{abc} \left( \frac{\lambda_7}{m^4} f_{cde} + \frac{\lambda_8}{m^4} d_{cde} \right) J^\alpha_{b} G^{\nu \beta} G_{\beta \alpha}
% \end{align}
\begin{align}\nonumber
       &  \frac{\lambda^2}{\lambda_R^2} \mathcal{D}_\mu G^{\mu \nu}_a + \frac{\lambda_1}{m^2} \mathcal{D}^\mu \left( f_{abc} G_{\mu \rho b} G^{\rho \nu}_c \right) - 8 \left[ \frac{\lambda_3}{m^4} k_{abcd} + \frac{\lambda_5}{m^4} f_{ace} f_{dbe} \right] \mathcal{D}_\mu \left( G^{\mu \nu}_b G_{\alpha \beta c} G^{\alpha \beta}_d \right)\\\nonumber
        & - 4 \left( \frac{\lambda_4}{m^4}f_{abe}f_{cde} + \frac{\lambda_6}{m^4} k_{abcd} \right) \mathcal{D}_\mu \left( G^{\nu \alpha}_b G^{\mu \beta}_{c} G_{\beta \alpha d} - G^{\mu \alpha}_b G^{\nu \beta}_{c} G_{\beta \alpha d} \right)\\\nonumber
       =  & \; J^\nu_a
       + \left( -\frac{\lambda_7}{m^4} f_{abc} + \frac{\lambda_8}{m^4} d_{abc} \right) \mathcal{D}_\mu \left[ \left( \mathcal{D}^\mu J_{\alpha b} \right) G^{\nu \alpha}_c - \left( \mathcal{D}^\nu J_{ \alpha b } \right) G^{\mu \alpha}_c + \left( \mathcal{D}_\alpha J^{ \nu}_{ c } \right) G^{\alpha \mu}_b - \left( \mathcal{D}_\alpha J_{ \mu c } \right) G^{\alpha \nu}_b \right]
       \\
       \nonumber
        & + 2 \frac{\lambda_9}{m^4} f_{abc} \mathcal{D}_\mu \left( J^{\mu}_b J^\nu_c \right) 
        - 2\frac{\lambda_{10}}{m^4} gf_{abc} J^\mu_b \left( \mathcal{D}^\nu J_{\mu c} \right) - f_{abc} \left( \frac{\lambda_7}{m^4} f_{cde} + \frac{\lambda_8}{m^4} d_{cde} \right) J^\alpha_{b} G^{\nu \beta}_d G_{\beta \alpha e}
        .
\end{align}

Accordingly, the Abelian limit for the two inhomogeneous Maxwell equations will be given by,
\begin{align}\nonumber
     & \frac{\lambda^2}{\lambda_R^2} \boldsymbol{\nabla} \cdot \mathbf{E} - 64 \frac{\lambda_3}{m^4} \boldsymbol{\nabla} \cdot \left( \mathbf{E} \left\{ \mathbf{H} \cdot \mathbf{H} - \mathbf{E} \cdot \mathbf{E} \right\} \right) - 16 \frac{\lambda_6}{m^4} \boldsymbol{\nabla} \cdot \left(  -\mathbf{E} \left\{ \mathbf{E} \cdot \mathbf{E} + \mathbf{H} \cdot \mathbf{H} \right\} + \mathbf{E} \left\{ \mathbf{H} \cdot \mathbf{H} - \mathbf{E} \cdot \mathbf{E} \right\}  - \mathbf{H} \left\{ \mathbf{E} \cdot \mathbf{H} - \mathbf{H} \cdot \mathbf{E} \right\} \right) 
    \\
    \nonumber
    & =  \rho_a
    - 4\frac{\lambda_8}{m^4} \boldsymbol{\nabla} \cdot \left( \left[ \boldsymbol{\nabla} J^i \right] E^i - \left[ \partial_t \rho \right] \mathbf{E} - \left[ \partial_t \mathbf{j} \right]\times \mathbf{H} - \left[ \partial_t \rho \right]\mathbf{E} - \left[ \boldsymbol{\nabla} \rho \right] \times \mathbf{H} + \left[ \partial^i \mathbf{j} \right] E^i \right)
\end{align}
and
\begin{align}
    \nonumber
    &  \frac{\lambda^2}{\lambda_R^2} \left( \partial_t \mathbf{E} - \boldsymbol{\nabla} \times \mathbf{H} \right) - 64 \frac{\lambda_3}{m^4} \left( \partial_t \left[ \mathbf{E} \left\{ \mathbf{H} \cdot \mathbf{H} - \mathbf{E} \cdot \mathbf{H}\right\} \right] - \boldsymbol{\nabla} \times \left[ \mathbf{H} \left\{ \mathbf{H} \cdot \mathbf{H} - \mathbf{E} \cdot \mathbf{H}\right\} \right] \right) \\\nonumber
    & - 16  \frac{\lambda_6}{m^4} \left( \partial_t \left[ 2\mathbf{E} \left\{ \mathbf{H} \cdot \mathbf{H} - \mathbf{E} \cdot \mathbf{E} \right\} - \mathbf{H} \left\{ \mathbf{H} \cdot \mathbf{E} + \mathbf{H} \cdot \mathbf{E} \right\} \right] \right. + \partial^j \left[ - 2\mathbf{E} \left\{ \mathbf{E} \times \mathbf{H}  \right\}^j + 2E^j  \mathbf{E} \times \mathbf{H} + \mathbf{H} \times \mathbf{H} H^j \right.\\\nonumber
    &\left. + \left\{ \mathbf{H} \times \mathbf{H} \right\}^j \mathbf{H} \right] \left. - 2 \boldsymbol{\nabla} \times \left[ \mathbf{H} \left\{ \mathbf{H} \cdot \mathbf{H} \right\} \right] \right) 
    \nonumber
    \\
     & =  \mathbf{j}
     + 4\frac{\lambda_8}{m^4} \left( \partial_t \left[ - \left( \partial_t  \mathbf{j} \right) \times \mathbf{H} + \left( \boldsymbol{\nabla} J^j \right) E^j + \left( \partial^j \mathbf{j} \right) E^j \right. \right. \left. + \left( \boldsymbol{\nabla} \rho \times \mathbf{H} \right) \right]\\\nonumber
    &\left. + \partial^j \left[ \left( \partial^j \rho \right) \mathbf{E} + \left( \partial^j \mathbf{j} \right) \times \mathbf{H} - \left( \boldsymbol{\nabla} \rho \right) E^j - \epsilon^{jkl} \left( \boldsymbol{\nabla} J^k \right) H^l + \left( \partial_t \mathbf{j} \right) E^j \right. \right. \left. \left. + \epsilon^{jkl} \left( \partial^k \mathbf{j} \right) H^l - \left( \partial_t J^j \right) \mathbf{E} - \left( \boldsymbol{\nabla} J^j \right) \times \mathbf{H} \right] \right)
\end{align}
Note that both equations receive several contributions from the imaginary term from the 
time-dependent and spatially non-homogeneous fermion-photon interactions.
They  indicate processes such as electromagnetic wave (or other configurations) emission or absorption and pair creation.

\end{document}